\documentclass[english,12pt,draftclsnofoot,english,onecolumn]{IEEEtran}
\usepackage[T1]{fontenc}
\usepackage{color}
\usepackage{babel}
\usepackage{amsmath}
\usepackage{amssymb}
\usepackage{stackrel}
\usepackage{graphicx}
\usepackage{setspace}
\usepackage[unicode=true,
 bookmarks=true,bookmarksnumbered=true,bookmarksopen=true,bookmarksopenlevel=1,
 breaklinks=false,pdfborder={0 0 0},pdfborderstyle={},backref=false,colorlinks=false]
 {hyperref}
\hypersetup{pdftitle={Your Title},
 pdfauthor={Your Name},
 pdfpagelayout=OneColumn, pdfnewwindow=true, pdfstartview=XYZ, plainpages=false}
\usepackage{breakurl}

\makeatletter
\ifCLASSOPTIONcompsoc
\usepackage[caption=false,font=normalsize,labelfont=sf,textfont=sf]{subfig}
\else
\usepackage[caption=false,font=footnotesize]{subfig}
\fi
\usepackage{cite}
\usepackage{amsthm}
\newtheorem{theorem}{\textbf{Theorem}}
\newtheorem{corollary}{\textbf{Corollary}}
\newtheorem{lemma}{\textbf{Lemma}}
\newtheorem{proposition}{\textbf{Proposition}}
\newtheorem{definition}{\textbf{Definition}}
\newtheorem{remark}{\textbf{Remark}}
\usepackage{algorithm}
\usepackage{amsmath}

\@ifundefined{showcaptionsetup}{}{%
 \PassOptionsToPackage{caption=false}{subfig}}
\usepackage{subfig}
\makeatother

\begin{document}

\title{MISO in Ultra-Dense Networks: Balancing the Tradeoff between User
and System Performance}

\author{\begin{small}\IEEEauthorblockN{Junyu Liu, Min Sheng, Jiandong
Li}\\
\IEEEauthorblockA{State Key Laboratory of Integrated Service Networks,
Xidian University, Xi'an, Shaanxi, 710071, China\\
Email: junyuliu@xidian.edu.cn, \{msheng, jdli\}@mail.xidian.edu.cn}\end{small}}
\maketitle
\begin{abstract}
With over-deployed network infrastructures, network densification
is shown to hinder the improvement of user experience and system performance.
In this paper, we adopt multi-antenna techniques to overcome the bottleneck
and investigate the performance of single-user beamforming, an effective
method to enhance desired signal power, in small cell networks from
the perspective of user coverage probability (CP) and network spatial
throughput (ST). Pessimistically, it is proved that, even when multi-antenna
techniques are applied, both CP and ST would be degraded and even
asymptotically diminish to zero with the increasing base station (BS)
density. Moreover, the results also reveal that the increase of ST
is at the expense of the degradation of CP. Therefore, to balance
the tradeoff between user and system performance, we further study
the critical density, under which ST could be maximized under the
CP constraint. Accordingly, the impact of key system parameters on
critical density is quantified via the derived closed-form expression.
Especially, the critical density is shown to be inversely proportional
to the square of antenna height difference between BSs and users.
Meanwhile, single-user beamforming, albeit incapable of improving
CP and ST scaling laws, is shown to significantly increase the critical
density, compared to the single-antenna regime.
\end{abstract}

\section{Introduction\label{sec:Introduction}}

Among the appealing approaches to fulfill the unprecedented capacity
goals of the future wireless networks, network densification is shown
to be the one with the greatest potential \cite{Network_densification_Ref1}.
The basic principle behind network densification is to deploy base
stations (BSs) or access points (APs) with smaller coverage to enable
local spectrum reuse \cite{Network_densification_Ref2,Network_densification_Ref3}.
As such, mobile users are served with short-distance transmission
links, thereby facilitating enormous spectrum reuse gain and enhancing
network capacity. The benefits of network densification are substantially
verified via analytical results from academia \cite{UPM_Ref_Original,UPM_Ref_K_tier,MUPM_Ref_Original,MUPM_Ref_LOS_Journal}
and experimental results from industries \cite{UDN_benefit_Qualcomm,UDN_benefit_ref}.
Remarkably, it is shown that over 1000-fold network capacity gain
can be harvested by deploying hundreds of self-organizing small cells
into one macro-cell, as compared to the macro-only case \cite{UDN_benefit_Qualcomm}.
Despite the merits, however, the results show that network capacity
starts to diminish when the number of small cells is sufficiently
large in ultra-dense networks (UDN) \cite{MUPM_Ref_Original,MUPM_Ref_LOS_Journal},
in which short-distance transmissions are more likely to occur and
accordingly inter-cell interference dominates the system performance.
For this reason, the limitation of network densification remains to
be fully explored.

\subsection{Related Work}

The research on how network densification impacts the capacity of
wireless networks has received extensive attention in the literature.
In \cite{UPM_Ref_Original,UPM_Ref_K_tier}, inspiring results have
been obtained, showing that network capacity can be sustainably increased
through deploying sufficient number of BSs in both single- and multi-tier
networks. Nevertheless, the analysis in \cite{UPM_Ref_Original,UPM_Ref_K_tier}
is made based on the premise that only non-line-of-sight (NLOS) paths
exist between the transmitters (Tx's) and the intended receivers (Rx's).
Due to the shorter transmission distance in dense BS deployment, line-of-sight
(LOS) paths are more likely to appear as well. On this account, authors
in \cite{MUPM_Ref_LOS_NLOS,MUPM_Ref_Original,MUPM_Ref_LOS_Journal}
made attempts to investigate the impact of LOS/NLOS transmissions
on the performance of downlink cellular networks. Particularly, it
has been reported that the user coverage probability (CP) is degraded
by network over-densification and, more importantly, network spatial
throughput (ST) grows sublinearly or even decreases with the growing
BS density \cite{MUPM_Ref_Original,MUPM_Ref_LOS_Journal,Ref_SBPM}.

In the aforementioned research, the 2-D distance is used to approximate
the distance between the antennas of Tx's and Rx's. In sparsely deployed
networks where Tx's and Rx's are far from each other, such approximation
is of high accuracy and thus valid. When Tx's and Rx's are in proximity,
however, it is apparent that the approximation will lose accuracy.
Considering more practical cases, authors in \cite{MUPM_Ref_K_dimension}
have investigated the performance of UDN in 3-D scenarios. Meanwhile,
the impact of antenna height difference (AHD) between Tx's and Rx's
has been examined in \cite{Ref_AHD_Ref1,Ref_AHD_Ref2}. In particular,
the results in \cite{Ref_AHD_Ref1} indicate that, considering the
existence of AHD, network capacity would even diminish to be zero
when the density of deployed BSs in UDN approaches infinity. Nevertheless,
as the obtained results are in complicated form, it fails to directly
characterize how network performance is affected by AHD under a reasonable
BS deployment density.

Intuitively, the main contributing factor that ruins the benefits
of network densification is the inter-cell interference, which is
likely to overwhelm the desired signal power when LOS paths exist
between interfering BSs and the intended downlink user. Especially,
when the density of elevated BSs further increases, more interfering
BSs would have LOS paths to the intended user, thereby degrading network
capacity. In this light, how to enhance desired signal power and mitigate
interference is of utmost importance in UDN. Recently, a number of
efficient interference management approaches have been tailored to
tackle the interference in UDN \cite{UDN_IM_Ref_clustering,UDN_IM_Ref_IA_RA,UDN_IM_Ref_Interference_Aware}.
For instance, an interference-separation clustering scheme has been
designed for UDN in \cite{UDN_IM_Ref_clustering}, aided by which
inter-cluster interference could be effectively avoided through BS
coordination. Besides, the combination of resource allocation and
interference alignment has been studied to mitigate interference in
UDN \cite{UDN_IM_Ref_IA_RA}. Nevertheless, the complexity of these
methods cannot be kept at a reasonably low level, since most of them
are run in a centralized manner. Worsestill, the required overhead
to implement these methods would unboundedly increase with the network
density, which may conversely ruin their potential benefits. Instead
of alleviating overwhelming interference, increasing the desired signal
power may serve as a promising alternative to enhance the system performance
in UDN as well. For instance, authors in \cite{Ref_UDN_JT} evaluate
the performance of cooperative transmissions in UDN. However, the
results indicate that user spectral efficiency can hardly be improved
by non-coherent joint transmission, whereas the spectral efficiency
gain brought by coherent joint transmission is considerably dependent
on channel models and system parameter settings. Therefore, more effective
schemes are to be developed to enhance the performance of UDN.

In addition to the above discussion, it should be noted that more
attentions have been paid on evaluating and improving the system-wide
performance of UDN in most of the existing researches \cite{MUPM_Ref_LOS_NLOS,MUPM_Ref_Original,MUPM_Ref_LOS_Conf,Ref_SBPM,Ref_AHD_Ref1,Ref_AHD_Ref2}.
Evidently, the quality of service (QoS) of users is an important indicator
to the performance of UDN as well. Nonetheless, it is shown in \cite{MUPM_Ref_Original}
that the user CP could only reach 0.2 in UDN under 10dB decoding threshold
and would even decrease with the growing BS density, which is typically
deemed unacceptable in practice. Moreover, it is shown from \cite{MUPM_Ref_Original}
that the improvement of system performance (e.g., network ST) is at
the cost of the deterioration of user performance (e.g., CP). For
the above reason, it is crucial to improve the QoS of users and balance
the tradeoff between user and network performance.

\subsection{Outcomes and Main Contribution}

In this paper, using the tools of stochastic geometry, we investigate
the performance of small cell networks, in which multiple antennas
are equipped on each elevated BS and single-user beamforming (SU-BF)
is applied as the multi-antenna technique. In particular, we provide
a tractable approach to analyze the CP (user performance) and ST (system
performance) in the multi-antenna regime, considering the AHD between
BSs and users. On this basis, the fundamental limitation of network
densification could be revealed. The main contribution of this paper
are summarized in the following:
\begin{itemize}
\item \textbf{Impact of AHD on the performance of UDN.} Considering the
antennas of BSs and users are of different heights, both CP and ST
are shown to be degraded by network over-densification. Meanwhile,
besides capturing the influence of AHD in the $\lambda\rightarrow\infty$
regime \cite{Ref_AHD_Ref1}, where $\lambda$ denotes the BS density,
we quantify the impact of AHD on CP and ST. In particular, it is revealed
that CP and ST would be exponentially decreased with the square of
AHD under typical settings.
\item \textbf{CP and ST scaling laws under the multi-antenna case.} We shed
light on the essential influence of SU-BF on the performance of UDN
by studying the CP and ST scaling laws. Notably, it is shown that
$\mathrm{\mathsf{CP}}\sim e^{-\bar{\kappa}\lambda}$ and $\mathsf{ST}\sim\lambda e^{-\bar{\kappa}\lambda}$,
where $\bar{\kappa}$ is a function of system parameters (excluding
BS density). In other words, even when multi-antenna techniques are
applied, network over-densification would totally drain the spectrum
reuse gain and degrade both user and system performance.
\item \textbf{Balancing the tradeoff between user and system performance.}
While SU-BF fails to improve the CP and ST scaling behavior, we show
that CP and ST could be significantly enhanced by SU-BF. More importantly,
to guarantee the QoS of users, we further analyze the critical density
that could maximize the network ST under the CP constraint. Specifically,
closed-form expressions of the critical density are retrieved in typical
cases, which capture the impact of key system parameters on critical
density. The above results could provide helpful insights and guidelines
towards the planning and deployment of future wireless networks.
\end{itemize}

For the remainder of this paper, we first present the system model
in Section \ref{sec:System-Model}, followed by a preliminary analysis
under the single-antenna regime in Section \ref{sec:Preliminary Analysis}.
Afterward, we investigate the performance of UDN when SU-BF is applied
in Section \ref{sec:analysis under SU-BF}, based on which the tradeoff
study on CP and ST is performed. Finally, conclusion remarks are given
in Section \ref{sec:Conclusion}.

\section{System Model\label{sec:System-Model}}

\subsection{Network Model\label{subsec:Network Model}}

\begin{figure}[t]
\begin{centering}
\includegraphics[width=3.5in]{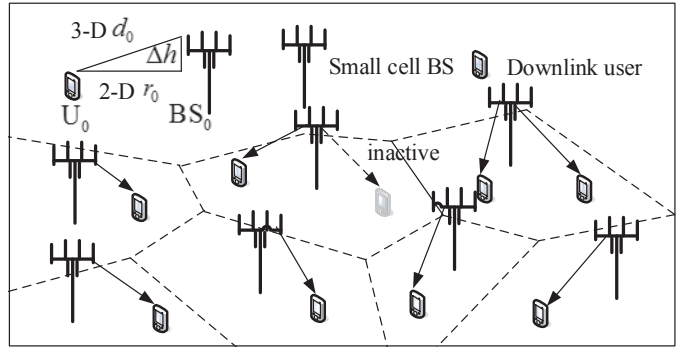}
\par\end{centering}
\caption{\label{fig:scenario}Illustration of downlink small cell networks.}
\end{figure}

Consider a\textcolor{black}{{} downlink small cell network (see Fig.
\ref{fig:scenario}), where BSs (with constant transmit power $P$)
and downlink users are distributed in a two-dimension plane $\mathbb{R}^{2}$,
in line with two independent homogeneous Poisson Point Processes (PPPs),
$\Pi_{\mathrm{BS}}=\left\{ \mathrm{BS}_{i}\left|\mathrm{BS}_{i}\in\mathbb{R}^{2}\right.\right\} $
and $\Pi_{\mathrm{U}}=\left\{ \mathrm{U}_{j}\left|\mathrm{U}_{j}\in\mathbb{R}^{2}\right.\right\} $
$\left(i,\:j\in\mathbb{N}\right)$, respectively. Denote $\lambda$
and $\lambda_{\mathrm{U}}$ as the densities of BSs and downlink users,
respectively. It is assumed that each multi-antenna BS is equipped
with antennas of height $h_{\mathrm{T}}$, while each single-antenna
downlink user is equipped with antenna of height $h_{\mathrm{R}}$.
Denote $\Delta h=\left|h_{\mathrm{T}}-h_{\mathrm{R}}\right|>0$ as
the AHD between BSs and users and $N_{\mathrm{a}}$ as the number
of antennas equipped on each BS. }

\textcolor{black}{The nearest association rule is adopted, i.e., downlink
users are associated with the geometrically nearest BSs. It is further
assumed that the user density is sufficiently large such that all
the BSs are connected and activated. In each time slot, each BS would
randomly select one of the associated users to serve. Besides, a saturated
data model is considered such that users always require data to download
from the serving BSs.}

\subsection{Single-user Beamforming\label{subsec:SU-BF}}

Instead of multi-user beamforming, SU-BF is applied as the multi-antenna
technique at each BS side to enhance user and system performance in
dense small cell networks. The main reasons are explained as follows.
For one BS, serving all connected users using multi-user beamforming
would require accurate estimation of channel state information (CSI)
from all served users, whereas imperfect CSI estimation would result
in significant performance degradation. In contrast, SU-BF, which
only requires the CSI from single user, is more favorable. Besides,
coordination among adjacent BSs is not considered, since it is difficult
to form the coordination cluster in UDN. Even when coordination cluster
is determined, CSI estimation and exchange within the coordination
cluster requires excessive overhead.

According to the above discussion, we denote the channel vector from
$\mathrm{BS}_{i}$ to $\mathrm{U}_{j}$ as $\mathbf{h}_{\mathrm{U}_{j},\mathrm{BS}_{i}}=\left[h_{\mathrm{U}_{j},\mathrm{BS}_{i},1},h_{\mathrm{U}_{j},\mathrm{BS}_{i},2},\ldots,h_{\mathrm{U}_{j},\mathrm{BS}_{i},N_{\mathrm{a}}}\right]$
with each complex entry independently distributed as complex normal
distribution with zero mean, i.e., $\mathcal{CN}\left(0,1\right)$,
and denote the SU-BF precoder from $\mathrm{BS}_{i}$ to $\mathrm{U}_{j}$
as $\mathbf{v}_{\mathrm{U}_{j},\mathrm{BS}_{i}}$, which is a unit
$1\times N_{\mathrm{a}}$ vector. If $s_{i}$ is the data symbol sent
by $\mathrm{BS}_{i}$, the received signal of $\mathrm{U}_{0}$, which
is served by $\mathrm{BS}_{0}$, is given by\footnote{Without loss of generality, the performance of the typical pair $\mathrm{BS}_{0}$-$\mathrm{U}_{0}$
is considered. Following\textcolor{black}{{} Slivnyak's Theorem \cite{book_stochastic_geometry}},
the performance of other pairs could be reflected by that of the typical
pair.}
\begin{align}
y_{0}= & s_{0}\mathbf{h}_{\mathrm{U}_{0},\mathrm{BS}_{0}}\mathbf{v}_{\mathrm{U}_{0},\mathrm{BS}_{0}}^{\mathrm{T}}l_{N}^{\frac{1}{2}}\left(\left\{ \alpha_{n}\right\} _{n=0}^{N-1};d_{0}\right)+\underset{\tiny{\mathrm{BS}_{i}\in\tilde{\Pi}_{\mathrm{BS}}}}{\sum}s_{i}\mathbf{h}_{\mathrm{U}_{0},\mathrm{BS}_{i}}\mathbf{v}_{\mathrm{U}_{0},\mathrm{BS}_{i}}^{\mathrm{T}}l_{N}^{\frac{1}{2}}\left(\left\{ \alpha_{n}\right\} _{n=0}^{N-1};d_{i}\right)+n_{0},\label{eq:received signal}
\end{align}
where $l_{N}\left(\left\{ \alpha_{n}\right\} _{n=0}^{N-1};d_{i}\right)$
denotes the pathloss from $\mathrm{BS}_{i}$ to $\mathrm{U}_{0}$,
$n_{0}$ denotes the additive Gaussian noise and $\tilde{\Pi}_{\mathrm{BS}}=\Pi_{\mathrm{BS}}\backslash\mathrm{BS}_{0}$.
In (\ref{eq:received signal}), $d_{i}$ denotes the distance from
the antenna of $\mathrm{BS}_{i}$ to that of $\mathrm{U}_{0}$ for
notation simplicity. Therefore, if denoting $\left\Vert \mathrm{BS}_{i}-\mathrm{U}_{0}\right\Vert $
as the 2D distance from $\mathrm{BS}_{i}$ to $\mathrm{U}_{0}$, we
have $d_{i}=\sqrt{\left\Vert \mathrm{BS}_{i}-\mathrm{U}_{0}\right\Vert ^{2}+\Delta h^{2}}$,
where the notation $\left\Vert \cdot\right\Vert $ denotes the Euclidean
norm operation. The detail of $l_{N}\left(\left\{ \alpha_{n}\right\} _{n=0}^{N-1};d_{i}\right)$
will be discussed later.

Assume that the CSI of the $\mathrm{BS}_{i}$-$\mathrm{U}_{i}$ pair
could be accurately estimated. In consequence, applying SU-BF would
contribute to $\left\Vert \mathbf{h}_{\mathrm{U}_{i},\mathrm{BS}_{i}}\mathbf{v}_{\mathrm{U}_{i},\mathrm{BS}_{i}}^{\mathrm{T}}\right\Vert \sim\chi_{2N_{\mathrm{a}}}^{2}$
and $\left\Vert \mathbf{h}_{\mathrm{U}_{i},\mathrm{BS}_{j}}\mathbf{v}_{\mathrm{U}_{i},\mathrm{BS}_{j}}^{\mathrm{T}}\right\Vert \sim\chi_{2}^{2}$
$\left(i\neq j\right)$ \cite{Ref_MIMO_2,Ref_MIMO_1}. For this reason,
the signal-to-interference ratio (SIR) at $\mathrm{U}_{0}$ can be
expressed as
\begin{align}
\mathsf{SIR}_{\mathrm{U}_{0}}=P\left\Vert \mathbf{h}_{\mathrm{U}_{0},\mathrm{BS}_{0}}\mathbf{v}_{\mathrm{U}_{0},\mathrm{BS}_{0}}^{\mathrm{T}}\right\Vert ^{2}l_{N}\left(\left\{ \alpha_{n}\right\} _{n=0}^{N-1};d_{0}\right)/I_{\mathrm{IC}} & ,\label{eq:SIR expression}
\end{align}
where $I_{\mathrm{IC}}=\underset{\tiny{\mathrm{BS}_{i}\in\tilde{\Pi}_{\mathrm{BS}}}}{\sum}P\left\Vert \mathbf{h}_{\mathrm{U}_{0},\mathrm{BS}_{i}}\mathbf{v}_{\mathrm{U}_{0},\mathrm{BS}_{i}}^{\mathrm{T}}\right\Vert ^{2}l_{N}\left(\left\{ \alpha_{n}\right\} _{n=0}^{N-1};d_{i}\right)$
denotes the intercell interference. It is worth noting that the influence
of noise on the user performance is neglected, as we consider the
interference-limited regime in UDN, where intercell interference dominates
the user and system performance.

\subsection{Pathloss Model}

To comprehensively characterize the LOS and NLOS components of signals
in UDN, a multi-slope pathloss model (MSPM) has been adopted as \cite{MUPM_Ref_Original,MUPM_Ref_LOS_Journal}
\begin{equation}
l_{N}\left(\left\{ \alpha_{n}\right\} _{n=0}^{N-1};x\right)=K_{n}x^{-\alpha_{n}},\:R_{n}\leq x<R_{n+1}\label{eq:MUPM}
\end{equation}
where $K_{0}=1$, $K_{n}=\prod_{i=1}^{n}R_{i}^{\alpha_{i}-\alpha_{i-1}}$
$\left(n\geq1\right)$, $0=R_{0}<R_{1}<\cdots<R_{N}=\infty$ and $0\leq\alpha_{0}\leq\alpha_{1}\leq\cdots\leq\alpha_{N-1}$
($\alpha_{N-1}>2$ for practical concerns \cite{MUPM_Ref_Original}).

From (\ref{eq:MUPM}), it follows that different pathloss exponents
are used to characterize the attenuation rates of signal power within
different regions. As a typical example, when $N=2$, MSPM degenerates
into the dual-slope pathloss model (DSPM) \cite{MUPM_Ref_Original,MUPM_Ref_K_dimension}
\begin{equation}
l_{2}\left(\left\{ \alpha_{n}\right\} _{n=0}^{1};x\right)=K_{n}x^{-\alpha_{n}},\:R_{n}\leq x<R_{n+1}\label{eq:DUPM}
\end{equation}
where $K_{0}=1$ and $K_{1}=R_{1}^{\alpha_{1}-\alpha_{0}}$. The DSPM
in (\ref{eq:DUPM}) is applied when an LOS path and a ground-reflected
path exist between Tx and the intended Rx. As such, signal power attenuates
slowly (with rate $\alpha_{0}$) within a \textit{corner distance}
$R_{1}$, while attenuates much more quickly (with rate $\alpha_{1}$)
with distance out of $R_{1}$. When $N=1$, MSPM further degenerates
into the most widely used single-slope pathloss model (SSPM) \cite{UPM_Ref_Original,MUPM_Ref_K_dimension}
\begin{equation}
l_{1}\left(\alpha_{0};x\right)=x^{-\alpha_{0}},\:x\in\left[0,\infty\right).\label{eq:SUPM}
\end{equation}

\subsection{Performance Metrics\label{subsec:Performance Metrics}}

We adopt CP and ST to reflect user and system performance, respectively.
To be specific, following the SIR at $\mathrm{U}_{0}$ in (\ref{eq:SIR expression}),
CP is defined as
\begin{equation}
\mathsf{CP}\left(\lambda\right)=\mathbb{P}\left\{ \mathsf{SIR}_{\mathrm{U}_{0}}>\tau\right\} ,\label{eq: define CP}
\end{equation}
where $\tau$ denotes the decoding threshold. Based on CP in (\ref{eq: define CP}),
we further define network ST as \cite{MUPM_Ref_Original,MUPM_Ref_K_dimension}
\begin{equation}
\mathsf{ST}\left(\lambda\right)=\lambda\mathbb{P}\left\{ \mathsf{SIR}_{\mathrm{U}_{0}}>\tau\right\} \log_{2}\left(1+\tau\right),\:\left[\mathrm{bits}/\left(\mathrm{s\cdot Hz\cdot m^{2}}\right)\right]\label{eq: define ST}
\end{equation}
which could characterize the number of bits that are successfully
conveyed over unit time, frequency and area.

\textbf{Notation}: In the following, the notations $\mathsf{CP}_{N}^{\mathrm{S}}\left(\lambda\right)$
(resp. $\mathsf{ST}_{N}^{\mathrm{S}}\left(\lambda\right)$) and $\mathsf{CP}_{N}^{\mathrm{M}}\left(\lambda\right)$
(resp. $\mathsf{ST}_{N}^{\mathrm{M}}\left(\lambda\right)$) will be
used. The superscript 'S' denotes SISO system, while the superscript
'M' denotes MISO system. The subscript $N$ denotes the number of
slopes in MSPM. If $_{2}F_{1}\left(\cdot,\cdot,\cdot,\cdot\right)$
is defined as the standard Gaussian hypergeometric function, denote
$\omega_{1}\left(x,y\right)=$ $_{2}F_{1}\left(1,1-\frac{2}{y},2-\frac{2}{y},-x\right)$,
$\omega_{2}\left(x,y\right)=$ $_{2}F_{1}\left(1,\frac{2}{y},1+\frac{2}{y},-x\right)$
and $\delta\left(x,y\right)=\frac{2x\omega_{1}\left(x,y\right)}{y-2}$
in the rest of the paper. Besides, we use $l_{N}\left(x\right)$ as
a substitution of $l_{N}\left(\left\{ \alpha_{n}\right\} _{n=0}^{N-1};x\right)$
for notation simplicity.

\section{Preliminary Analysis\label{sec:Preliminary Analysis}}

In this section, we provide preliminary analysis of CP and ST under
MSPM when single antenna is equipped by each BS. The purpose is to
lay the foundation for the analysis of multi-antenna case in Section
\ref{sec:analysis under SU-BF}.

\subsection{CP and ST in SISO system}

When each BS is equipped with one antenna, no precoder is to be designed
and, accordingly, the SIR at $\mathrm{U}_{0}$ in (\ref{eq:SIR expression})
would degenerate into
\begin{align}
\mathsf{SIR}_{\mathrm{U}_{0}}^{\mathrm{S}}=P\left\Vert h_{\mathrm{U}_{0},\mathrm{BS}_{0}}\right\Vert ^{2}l_{N}\left(\left\{ \alpha_{n}\right\} _{n=0}^{N-1};d_{0}\right) & /I_{\mathrm{IC}}^{\mathrm{S}},\label{eq:SIR expression SISO}
\end{align}
where $I_{\mathrm{IC}}^{\mathrm{S}}=\underset{\tiny{\mathrm{BS}_{i}\in\tilde{\Pi}_{\mathrm{BS}}}}{\sum}P\left\Vert h_{\mathrm{U}_{0},\mathrm{BS}_{i}}\right\Vert ^{2}l_{N}\left(\left\{ \alpha_{n}\right\} _{n=0}^{N-1};d_{i}\right)$
denotes the intercell interference in the SISO system and $h_{\mathrm{U}_{0},\mathrm{BS}_{i}}$
denotes the channel from $\mathrm{BS}_{i}$ to $\mathrm{U}_{0}$.

In practice, when LOS path appears between Tx and the intended Rx,
$h_{\mathrm{U}_{0},\mathrm{BS}_{i}}$ is more likely to follow complex
normal distribution with non-zero mean (Rice fading), which is inconsistent
with the $h_{\mathrm{U}_{0},\mathrm{BS}_{i}}\sim\mathcal{CN}\left(0,1\right)$
(Rayleigh fading) assumption in Section \ref{subsec:SU-BF}. Nevertheless,
we have tested via the experiment that signal envelop still follows
Rayleigh distribution when Tx's and Rx's are geometrically close enough
(several meters to dozens of meters), since the signal strengths of
LOS and NLOS components are comparable \cite{Ref_Test_BPM}. More
importantly, as will be shown later, the assumption on small-scale
fading, which is considered as the minor factor to influence the performance
of UDN \cite{Ref_Minor_factor}, would exert little impact on CP and
ST scaling laws.

Following (\ref{eq:SIR expression SISO}), we give the main results
on CP and ST under MSPM in SISO systems in Proposition \ref{proposition: CP and ST SISO}.

\begin{proposition}

Considering the AHD between single-antenna BSs and downlink users,
the ST in downlink small cell networks under MSPM in (\ref{eq:MUPM})
is given by $\mathsf{ST}_{N}^{\mathrm{S}}\left(\lambda\right)=\lambda\mathsf{CP}_{N}^{\mathrm{S}}\left(\lambda\right)\log_{2}\left(1+\tau\right)$,
where $\mathsf{CP}_{N}^{\mathrm{S}}\left(\lambda\right)$ is given
by (\ref{eq:CP general SISO}).

\begin{equation}
\mathsf{CP}_{N}^{\mathrm{S}}\left(\lambda\right)=\left\{ \begin{array}{cc}
\frac{1}{1+\delta\left(\tau,\alpha_{0}\right)}\exp\left(-\pi\lambda\delta\left(\tau,\alpha_{0}\right)\Delta h^{2}\right), & N=1\\
\stackrel[n=0]{N-1}{\sum}\mathbb{E}_{r_{0}\in\left[R_{n},R_{n+1}\right)}\left\{ \exp\left[-\pi\lambda\left(\bar{R}_{n+1}^{2}\omega_{2}\left(\frac{\bar{R}_{n+1}^{\alpha_{n}}}{\tau d_{0}^{\alpha_{n}}},\alpha_{n}\right)-d_{0}^{2}\omega_{2}\left(\tau^{-1},\alpha_{n}\right)\right.\right.\right.\\
+\left.\left.\left.\stackrel[i=n+1]{N-1}{\sum}\left(\bar{R}_{i+1}^{2}\omega_{2}\left(\frac{\bar{R}_{i+1}^{\alpha_{i}}}{\tau K_{i}d_{0}^{\alpha_{n}}},\alpha_{i}\right)-\bar{R}_{i}^{2}\omega_{2}\left(\frac{\bar{R}_{i}^{\alpha_{i}}}{\tau K_{i}d_{0}^{\alpha_{n}}},\alpha_{i}\right)\right)\right)\right]\right\} , & N>1
\end{array}\right.\label{eq:CP general SISO}
\end{equation}

In (\ref{eq:CP general SISO}), $d_{0}=\sqrt{r_{0}^{2}+\Delta h^{2}}$
and the probability density function (PDF) of $r_{0}$ is derived
from the contact distribution \cite{book_stochastic_geometry}
\begin{equation}
f_{r_{0}}\left(x\right)=2\pi\lambda x\exp\left(-\pi\lambda x^{2}\right),\:x\geq0.\label{eq:PDF of r0}
\end{equation}

\label{proposition: CP and ST SISO}

\end{proposition}

\textit{Proof}: Please refer to Appendix \ref{subsec:Proof for SP and ST SISO}.\qed

Despite in complicated form, the results in Proposition \ref{proposition: CP and ST SISO}
could provide a numerical approach to capture the influence of system
parameters on CP and ST, under MSPM. Meanwhile, according to the special
case in (\ref{eq:CP general SISO}), where $N=1$, it follows that
both CP and ST would exponentially decrease with $\Delta h^{2}$.
In other words, if ignoring the impact of the AHD (i.e., $\Delta h=0$),
the merits of network densification would be greatly over-estimated.
In addition, when MSPM degenerates into DSPM, the results on CP and
ST could be further simplified as follows.

\begin{corollary}

Considering the AHD between single-antenna BSs and downlink users,
the ST in downlink small cell networks under DSPM in (\ref{eq:DUPM})
is given by $\mathsf{ST}_{2}^{\mathrm{S}}\left(\lambda\right)=\lambda\mathsf{CP}_{2}^{\mathrm{S}}\left(\lambda\right)\log_{2}\left(1+\tau\right)$,
where
\begin{equation}
\mathsf{CP}_{2}^{\mathrm{S}}\left(\lambda\right)=\mathbb{E}_{r_{0}\in\left[0,R_{1}\right)}\left[e^{-\pi\lambda\left(\delta_{1}\left(\alpha_{0},d_{0},\tau,R_{1}\right)+\delta_{2}\left(\alpha_{0},\alpha_{1},d_{0},\tau,R_{1}\right)\right)}\right]+\frac{e^{-\pi\lambda R_{1}^{2}\left(1+\delta\left(\tau,\alpha_{1}\right)\right)}}{1+\delta\left(\tau,\alpha_{1}\right)}.\label{eq:CP DUPM}
\end{equation}
In (\ref{eq:CP DUPM}), $d_{0}=\sqrt{r_{0}^{2}+\Delta h^{2}}$, $\delta_{1}\left(\alpha_{0},d_{0},\tau,R_{1}\right)=R_{1}^{2}\omega_{2}\left(\frac{R_{1}^{\alpha_{0}}}{\tau d_{0}^{\alpha_{0}}},\alpha_{0}\right)-d_{0}^{2}\omega_{2}\left(\frac{1}{\tau},\alpha_{0}\right)$,
$\delta_{2}\left(\alpha_{0},\alpha_{1},d_{0},\tau,R_{1}\right)=\frac{2\tau d_{0}^{\alpha_{0}}R_{1}^{2-\alpha_{0}}}{\alpha_{1}-2}\omega_{1}\left(\frac{\tau d_{0}^{\alpha_{0}}}{R_{1}^{\alpha_{0}}},\alpha_{1}\right)$,
$\delta\left(\tau,\alpha_{1}\right)=\frac{2\tau\omega_{1}\left(\tau,\alpha_{1}\right)}{\alpha_{1}-2}$
and the PDF of $r_{0}$ is given by (\ref{eq:PDF of r0}).

\label{corollary: CP and ST DUPM SISO}

\end{corollary}

\textit{Proof}: The proof can be completed by setting $N=2$ in (\ref{eq:CP general SISO})
with easy manipulation, and is thus omitted due to space limitation.\qed

\begin{figure}[t]
\begin{centering}
\subfloat[CP.]{\begin{centering}
\includegraphics[width=3in]{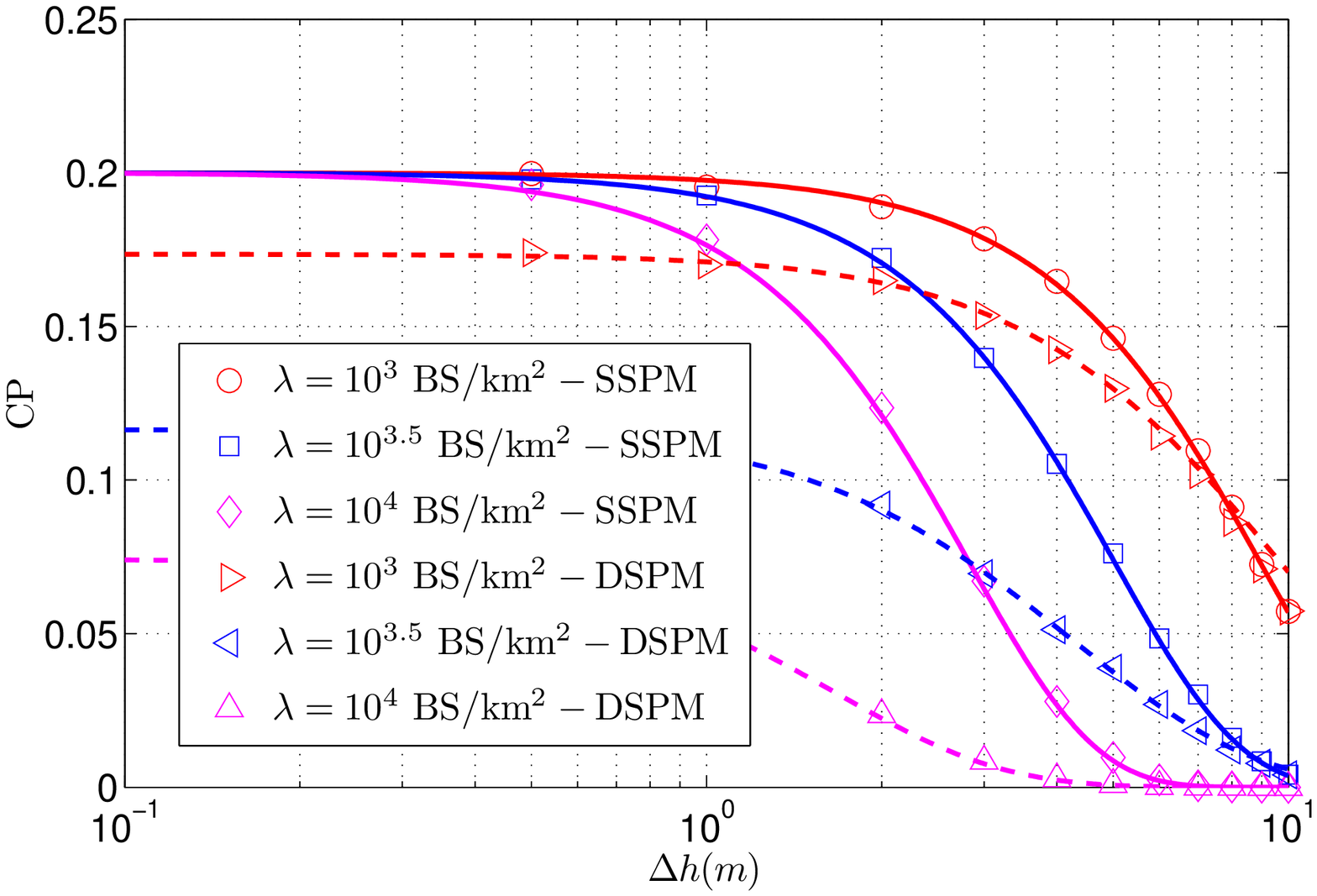}
\par\end{centering}
}\subfloat[ST.]{\begin{centering}
\includegraphics[width=3in]{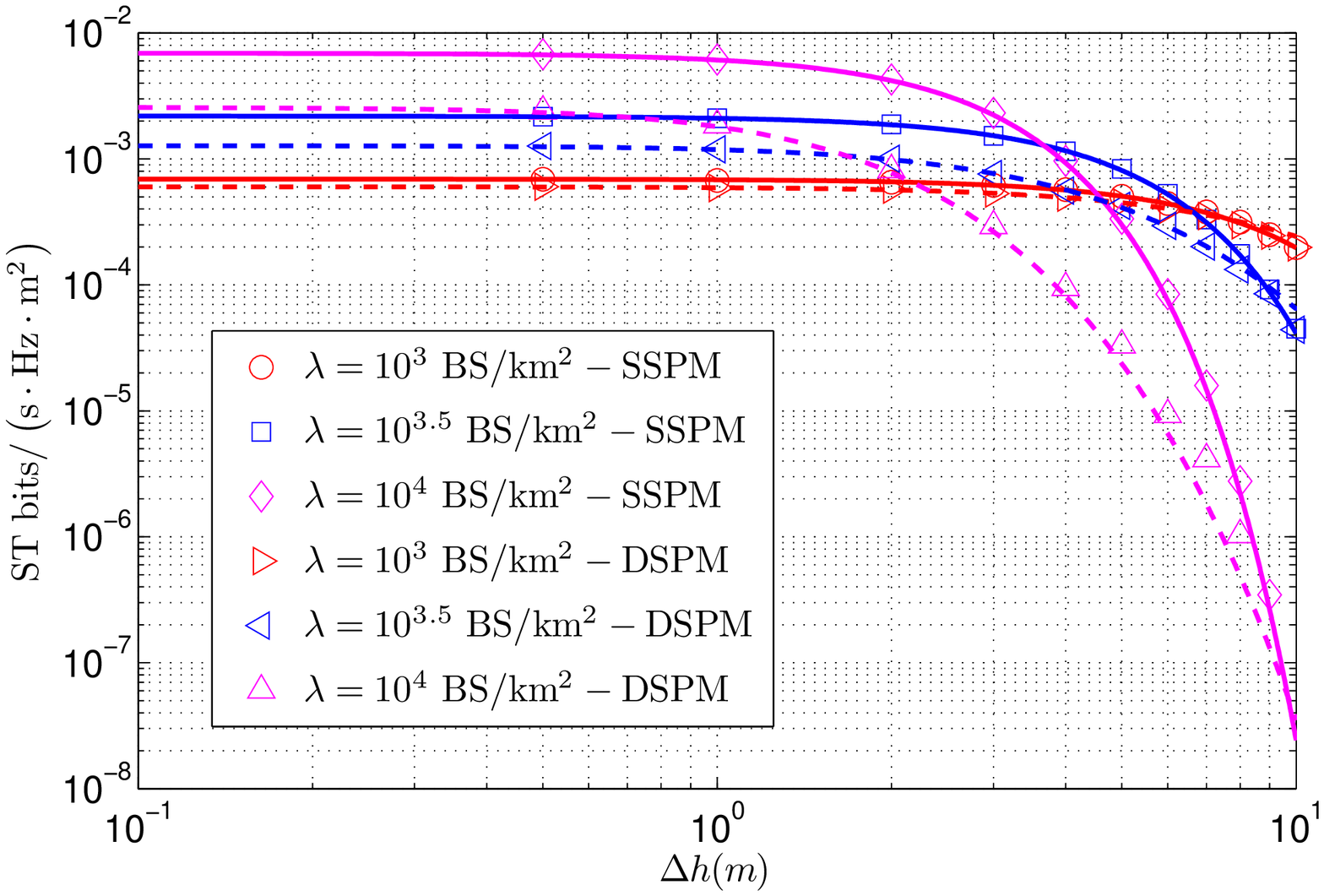}
\par\end{centering}
}
\par\end{centering}
\caption{\label{fig:CP and ST with AHD}CP and ST varying with AHD $\Delta h$.
For system settings, set $P=23$dBm and $\tau=10$dB. For SSPM, set
$\alpha_{0}=4$. For DSPM, set $\alpha_{0}=2.5$, $\alpha_{1}=4$
and $R_{1}=10$m. Lines and markers denote numerical and simulation
results, respectively, in this figure and the remaining figures in
this paper.}
\end{figure}

Based on Proposition \ref{proposition: CP and ST SISO} and Corollary
\ref{corollary: CP and ST DUPM SISO}, we illustrate the impact of
AHD on CP and network ST in detail. In particular, Fig. \ref{fig:CP and ST with AHD}
plots the CP and ST as a function of $\Delta h$ under different BS
densities. It is shown that both CP and ST would be degraded by $\Delta h$.
This indicates that, although the existence of $\Delta h$ would weaken
both desired and interference signal power, the decrease of the desired
signal power overwhelms that of the interference signal powers. Meanwhile,
it is shown that the impact of $\Delta h$ on CP and ST is significant
under dense BS deployment. Hence, in dense wireless networks, where
the user antenna heights are basically small, it is preferable to
deploy small cell BSs with smaller antenna heights so as to ensure
the user performance as well as system performance.

As shown in Fig. \ref{fig:CP and ST with AHD}, it is evident that
the existence of $\Delta h$ leads to the performance degradation
of downlink small cell networks in terms of CP and ST, especially
when BSs are densely deployed. Therefore, we intend to further explore
the influence of $\Delta h$ on the scaling laws of CP and ST in the
following.

\subsection{CP and ST scaling laws in SISO system}

In this part, before investigating the CP and ST scaling laws, results
on $\omega_{1}\left(x,y\right)$ are first given in the following
lemma.

\begin{lemma}

For $y>2$, $\omega_{1}\left(x,y\right)$ is a decreasing function
of $x$.

\label{lemma: hypergeometric function}

\end{lemma}

\textit{Proof}: Please refer to the proof for Lemma 1 in \cite{Ref_SBPM}.\qed

To study the CP and ST scaling laws, the definition, which describes
the growth rate (or order) of a function, is further given in the
following.

\begin{definition}[Function Limiting Behavior]

Denote $g_{1}\left(x\right)$ and $g_{2}\left(x\right)$ as two functions
defined on the subset of real numbers. We write $g_{1}\left(x\right)=\Omega\left(g_{2}\left(x\right)\right)$
if $\exists m>0$, $x_{0}$, $\forall x>x_{0}$, $m\left|g_{2}\left(x\right)\right|\leq\left|g_{1}\left(x\right)\right|$,
and $g_{1}\left(x\right)=\mathcal{O}\left(g_{2}\left(x\right)\right)$
if $\exists m>0$, $x_{0}$, $\forall x>x_{0}$, $\left|g_{1}\left(x\right)\right|\leq m\left|g_{2}\left(x\right)\right|$.

\label{definition: scaling behavior}

\end{definition}

Aided by Definition \ref{definition: scaling behavior} and Proposition
\ref{proposition: CP and ST SISO}, we show the CP and ST scaling
laws in Theorem \ref{theorem: CP and ST scaling law SISO}.

\begin{theorem}[CP and ST Scaling Laws in SISO System]

When AHD exists between single-antenna BSs and downlink users, CP
and ST scale with BS density $\lambda$ as $\mathsf{CP}_{N}^{\mathrm{S}}\left(\lambda\right)\sim e^{-\kappa\lambda}$
and $\mathsf{ST}_{N}^{\mathrm{S}}\left(\lambda\right)\sim\lambda e^{-\kappa\lambda}$
($\kappa$ is a constant), respectively.

\label{theorem: CP and ST scaling law SISO}

\end{theorem}

\textit{Proof}: Please refer to Appendix \ref{subsec:Proof for scaling law}.\qed

It is shown from Theorem \ref{theorem: CP and ST scaling law SISO}
that both user and system performance would be degraded when BS density
is sufficiently large. This is essentially different from the results
in \cite{MUPM_Ref_LOS_NLOS,MUPM_Ref_Original,MUPM_Ref_LOS_Conf,MUPM_Ref_LOS_Journal},
where BSs and users are equipped with antennas of the identical heights
and the impact of AHD has not been taken into account in the scaling
law analysis. Before showing the difference, we give the definition
on critical density to facilitate better understanding of ST scaling
law.

\begin{definition}[Critical density]

Critical density is defined as the BS density, which maximizes network
ST. Beyond the critical density, network ST starts to diminish with
the growing BS density.

\label{definition: critical density}

\end{definition}

The critical density in Definition \ref{definition: critical density}
serves as a useful metric to reflect the maximal density of BSs that
could be deployed without degrading network capacity. Therefore, it
can be used to reveal the fundamental limitation of network densification
under different system settings.

\begin{figure}[t]
\begin{centering}
\subfloat[\label{fig:CP scaling law}CP.]{\begin{centering}
\includegraphics[width=3in]{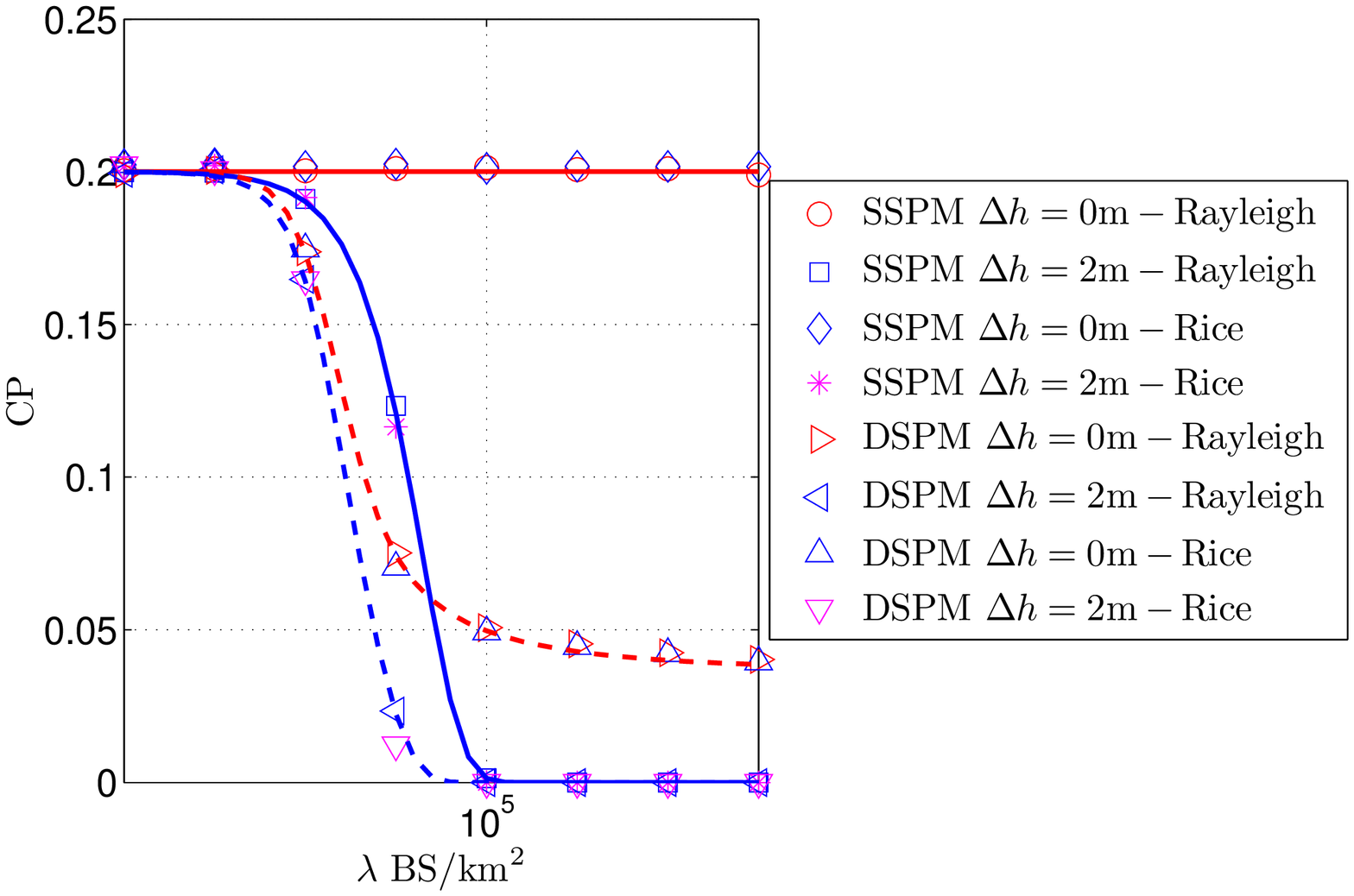}
\par\end{centering}
}\subfloat[\label{fig:ST scaling law}ST.]{\begin{centering}
\includegraphics[width=3in]{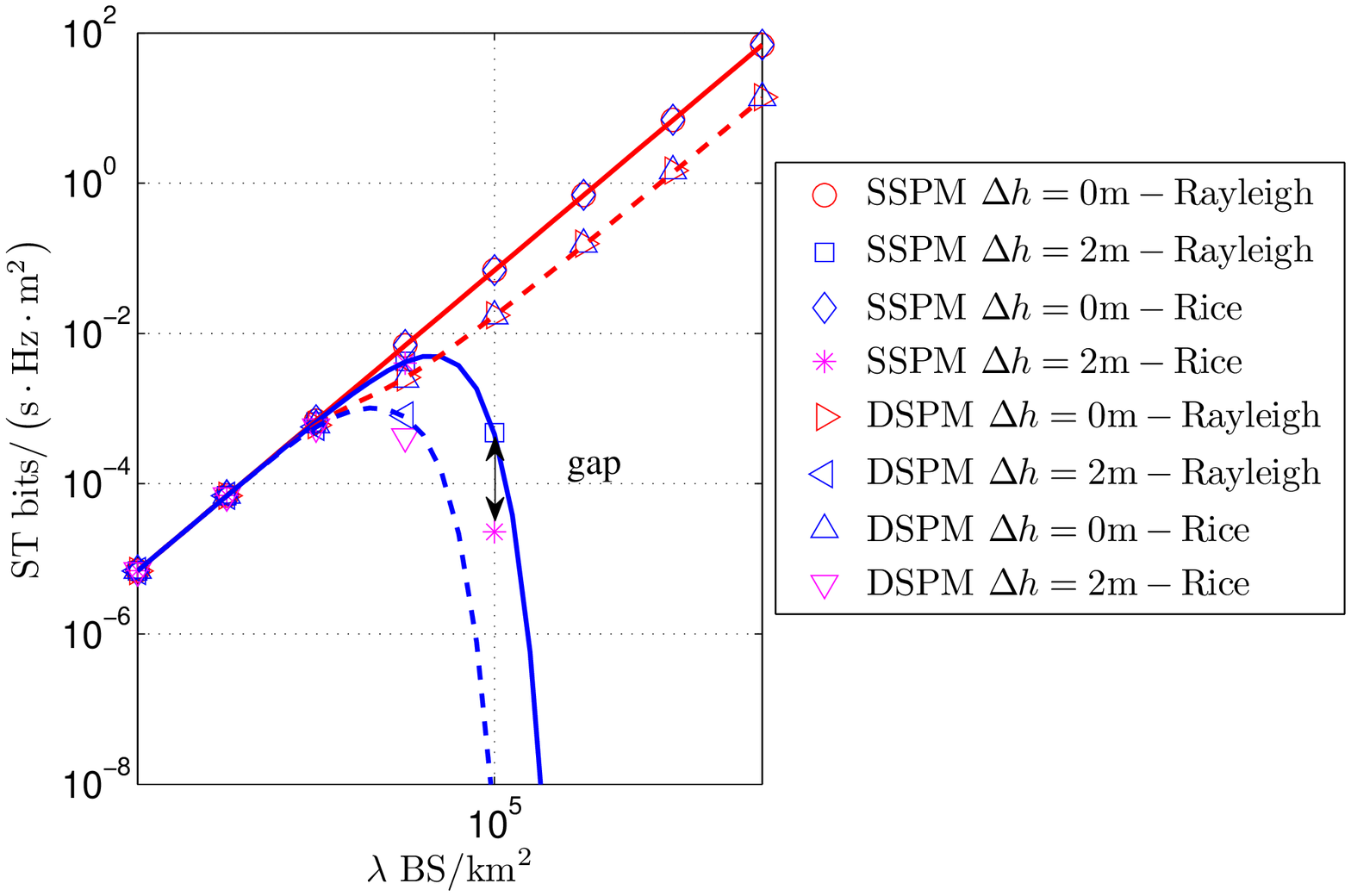}
\par\end{centering}
}
\par\end{centering}
\caption{\label{fig:CP and ST scaling law}CP and ST varying with BS density
$\lambda$. For system settings, set $P=23$dBm and $\tau=10$dB.
For SSPM, set $\alpha_{0}=4$. For DSPM, set $\alpha_{0}=2.5$, $\alpha_{1}=4$
and $R_{1}=10$m. To reflect the impact of LOS paths on signal propagation,
we set $\upsilon_{\mathrm{NC}}=1$ and $\upsilon_{\mathrm{DoF}}=12$
for Rice fading.}
\end{figure}

Fig. \ref{fig:CP and ST scaling law} shows the CP and ST as a function
of BS density $\lambda$ under different $\Delta h$. It is shown
in Fig. \ref{fig:CP scaling law} that, when $\Delta h=0$m, CP almost
keeps constant with the increasing $\lambda$ under SSPM, and slowly
decreases with the increasing $\lambda$ under DSPM (compared to the
DSPM case under $\Delta h>0$m). In consequence, network ST would
linearly/sublinearly grow with $\lambda$, as shown in Fig. \ref{fig:ST scaling law}.
In contrast, both CP and ST asymptotically approach zero when $\lambda$
is sufficiently large under $\Delta h>0$m. In practice, AHD exists
between BSs and downlink users, even when small cell BSs are densely
deployed. Therefore, the results in Theorem \ref{theorem: CP and ST scaling law SISO}
could shed light on the fundamental limitation of network densification.

To verify the validity of the scaling law analysis under Rayleigh
fading, we also evaluate the performance of downlink networks under
Rice fading via simulation results in Fig. \ref{fig:CP and ST scaling law}.
Specifically, the channel power gain under Rice fading channels follows
the non-central chi-square distribution with non-centrality parameter
$\upsilon_{\mathrm{NC}}$ and degrees of freedom $\upsilon_{\mathrm{DoF}}$.
It can be seen from Fig. \ref{fig:CP and ST scaling law} that, although
gaps exist between the results under Rice and Rayleigh fadings, it
is apparent that the CP and ST scaling laws under Rice fading are
identical as those under Rayleigh fading. Following the above analysis,
we further investigate the impact of SU-BF on CP and ST.

\section{Performance Analysis Under SU-BF\label{sec:analysis under SU-BF}}

In this section, we evaluate the performance of SU-BF in downlink
small cell networks especially when BSs are densely deployed. To this
end, we first extend the results on CP (resp. ST) and CP scaling law
(resp. ST scaling law) in the following.

\subsection{CP and ST in multi-antenna case}

When SU-BF is applied at the BS side, the SIR at the typical downlink
user $\mathrm{U}_{0}$ is given by (\ref{eq:SIR expression}) in Section
\ref{subsec:SU-BF}. Aided by the preliminary analysis in Section
\ref{sec:Preliminary Analysis}, CP and ST can be derived via easy
extension in the following corollary.

\begin{corollary}

When SU-BF is applied by each multi-antenna BS, the ST in downlink
small cell networks under MSPM in (\ref{eq:MUPM}) is given by $\mathsf{ST}_{N}^{\mathrm{M}}\left(\lambda\right)=\lambda\mathsf{CP}_{N}^{\mathrm{M}}\left(\lambda\right)\log_{2}\left(1+\tau\right)$,
where
\begin{align}
\mathsf{CP}_{N}^{\mathrm{M}}\left(\lambda\right)= & \mathbb{E}\left[\stackrel[k=0]{N_{\mathrm{a}}-1}{\sum}\frac{\left(-s\right)^{k}}{k!}\frac{\mathrm{d}^{k}}{\mathrm{d}s^{k}}\mathcal{L}_{I_{\mathrm{IC}}}\left(s\right)\left|_{s=\frac{\tau}{2Pl_{N}\left(d_{0}\right)}}\right.\right].\label{eq:CP general MISO}
\end{align}
In (\ref{eq:CP general MISO}), $\mathcal{L}_{I_{\mathrm{IC}}}\left(s\right)$
is the Laplace Transform of $I_{I_{\mathrm{IC}}}$ evaluated at $s=\frac{\tau}{2Pl_{N}\left(d_{0}\right)}$,
which is given by
\begin{align*}
\mathcal{L}_{I_{\mathrm{IC}}}\left(s\right)= & \exp\left(-2\pi\lambda\int_{d_{0}}^{\infty}x\left(1-\frac{1}{1+2sPl_{N}\left(x\right)}\right)\mathrm{d}x\right).
\end{align*}

\label{corollary: CP and ST MISO}

\end{corollary}

\textit{Proof}: Please refer to Appendix \ref{subsec:Proof for Corollary CP and ST MISO}.\qed

\begin{figure}[t]
\begin{centering}
\subfloat[\label{fig:CP MISO different TH}CP.]{\begin{centering}
\includegraphics[width=3in]{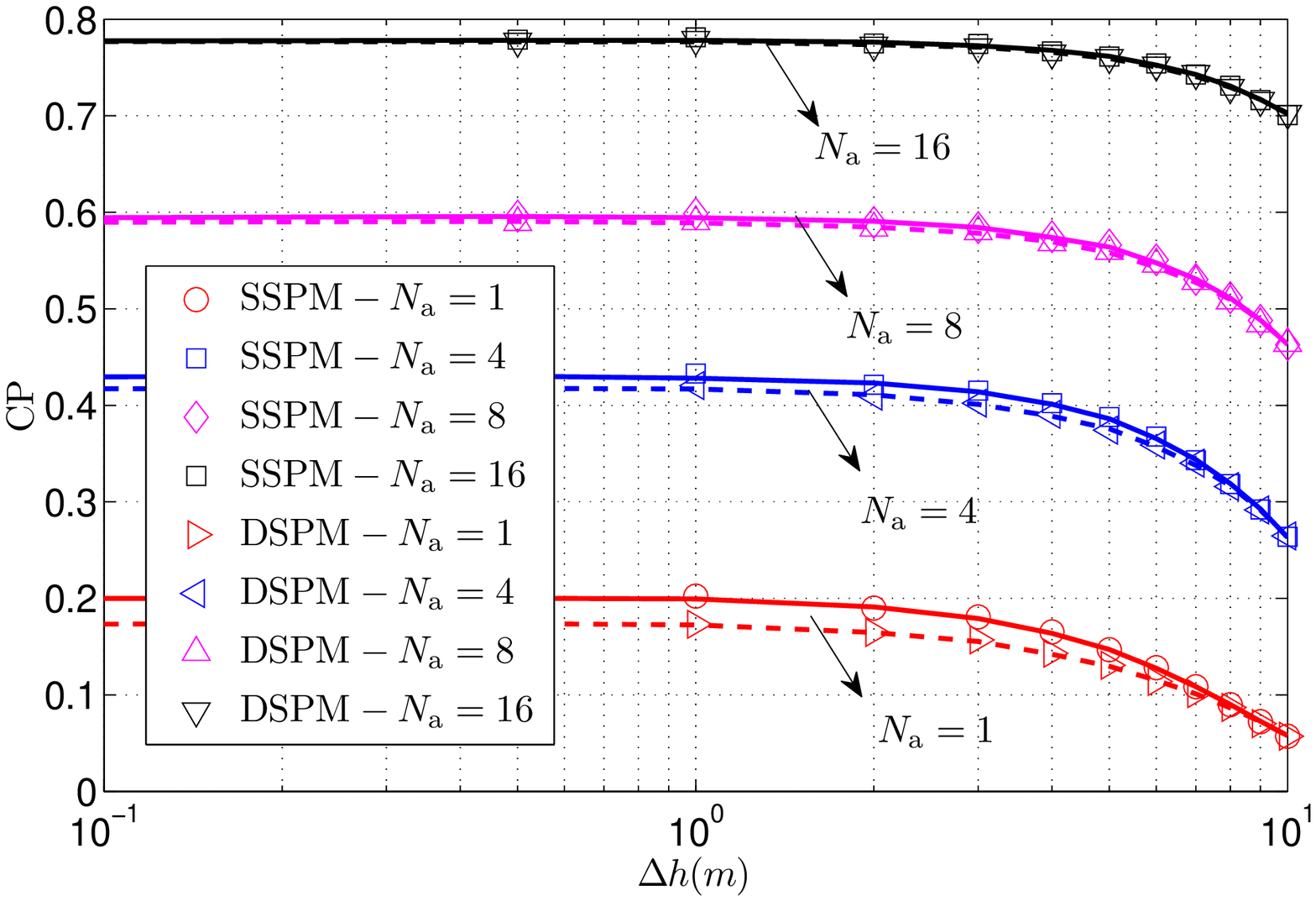}
\par\end{centering}
}\subfloat[\label{fig:ST MISO different TH}ST.]{\begin{centering}
\includegraphics[width=3in]{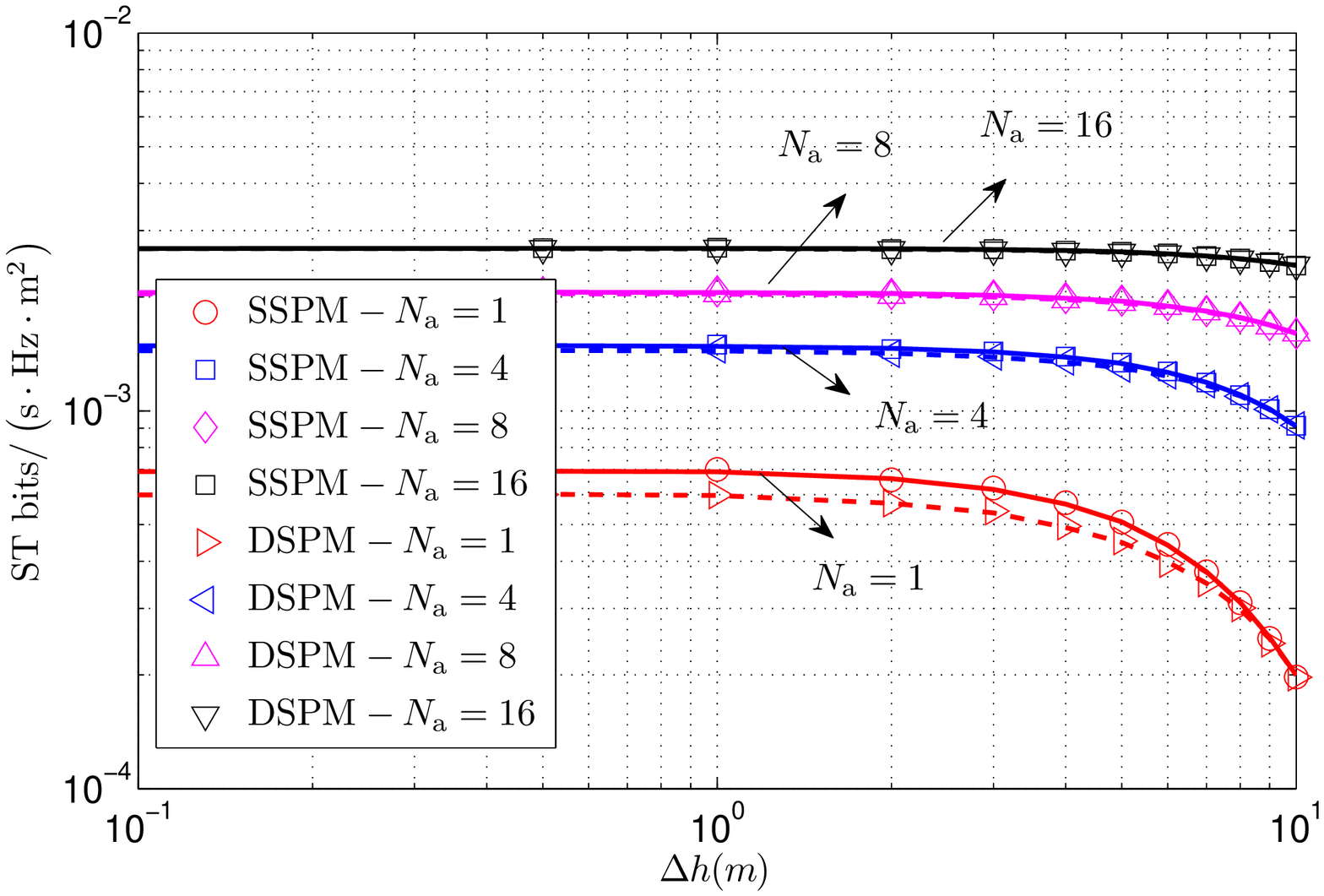}
\par\end{centering}
}
\par\end{centering}
\caption{\label{fig:CP and ST MISO different TH}CP and ST varying with AHD
$\Delta h$. For system settings, set $P=23$dBm and $\tau=10$dB.
For SSPM, set $\alpha_{0}=4$. For DSPM, set $\alpha_{0}=2.5$, $\alpha_{1}=4$
and $R_{1}=10$m.}
\end{figure}

With the aid of Corollary \ref{corollary: CP and ST MISO}, we illustrate
how SU-BF enhances user and system performance in Fig. \ref{fig:CP and ST MISO different TH}.
In particular, Fig. \ref{fig:CP and ST MISO different TH} plots the
CP and ST as a function of AHD $\Delta h$ when different number of
antennas are equipped on each BS. It is observed that CP and ST could
be greatly improved when SU-BF is applied. As an example, CP could
only reach 0.2 in the single-antenna case under $\Delta h=1$m in
Fig. \ref{fig:CP MISO different TH}. When SU-BF is used with $N_{\mathrm{a}}=16$,
however, CP could be increased to 0.78 under $\Delta h=1$m. Meanwhile,
we see from Fig. \ref{fig:CP and ST MISO different TH} that CP and
ST would be degraded more slowly with $\Delta h$ when more antennas
are equipped on each BS. This also confirms the benefits of SU-BF
in downlink small cell networks. In the following, we further investigate
the performance of SU-BF in dense scenarios by studying CP and ST
scaling laws.

\begin{theorem}[CP and ST Scaling Laws in MISO System]

When SU-BF is applied, CP and ST scale with BS density $\lambda$
as $\mathsf{CP}_{N}^{\mathrm{M}}\left(\lambda\right)\sim e^{-\bar{\kappa}\lambda}$
and $\mathsf{ST}_{N}^{\mathrm{M}}\left(\lambda\right)\sim\lambda e^{-\bar{\kappa}\lambda}$
($\bar{\kappa}$ is a constant), respectively, under MSPM.

\label{theorem: CP and ST scaling law MISO}

\end{theorem}

\textit{Proof}: Please refer to Appendix \ref{subsec:Proof for scaling law MISO}.\qed

\begin{figure*}[t]
\begin{centering}
\subfloat[\label{fig:CP scaling law MISO}CP.]{\begin{centering}
\includegraphics[width=2.1in]{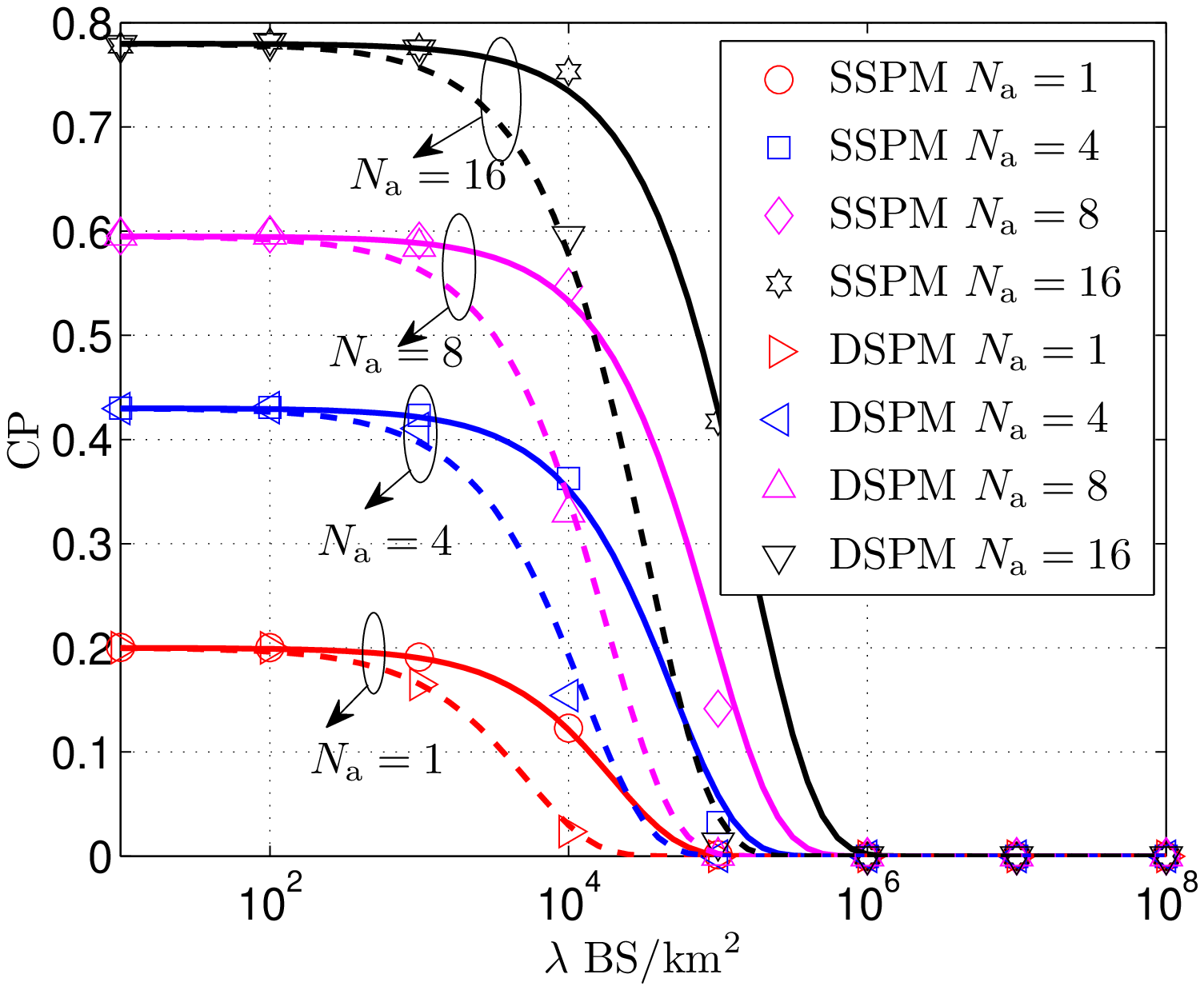}
\par\end{centering}
}\subfloat[\label{fig:ST scaling law MISO SSPM}ST under SSPM.]{\begin{centering}
\includegraphics[width=2.1in]{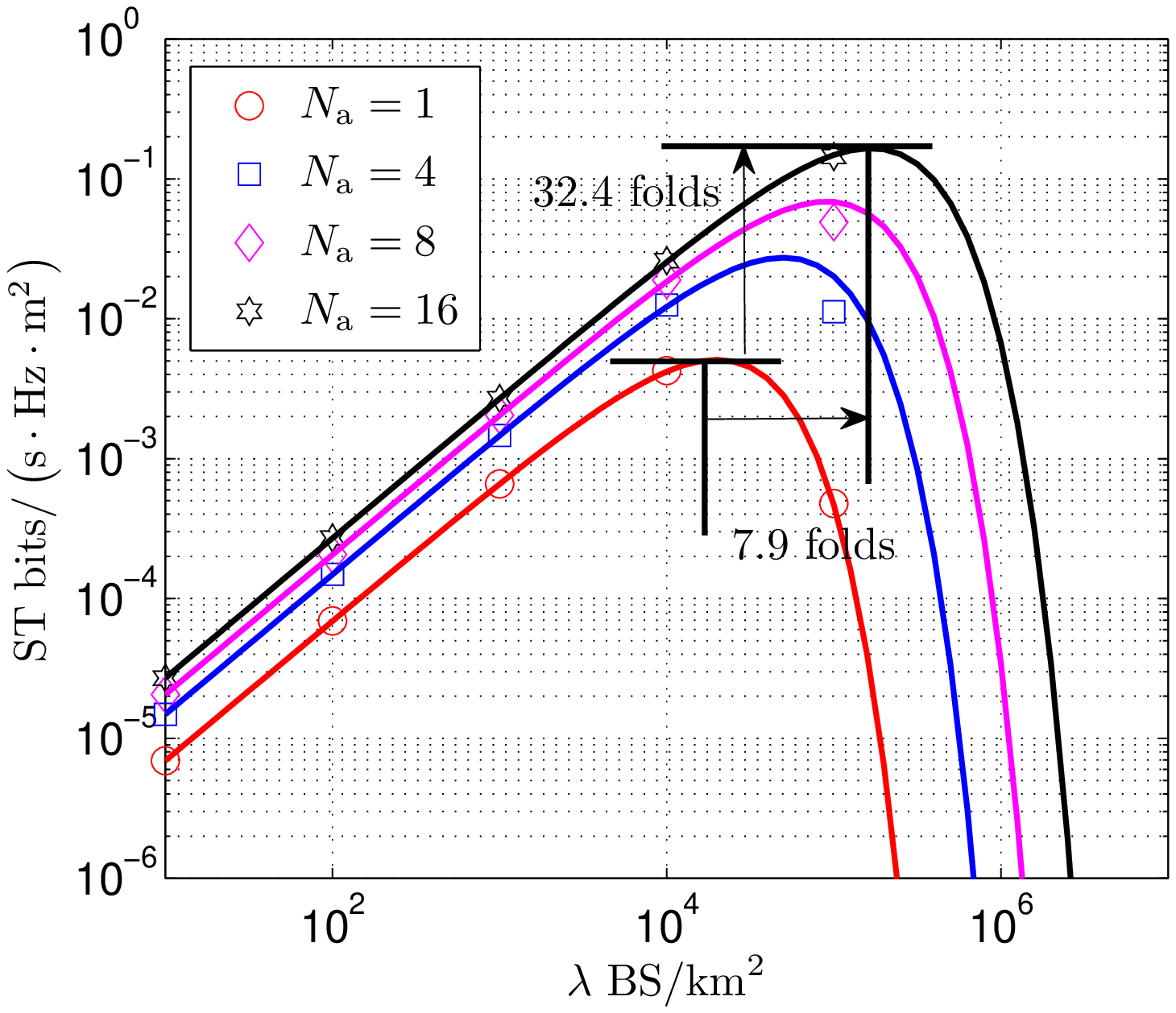}
\par\end{centering}
}\subfloat[\label{fig:ST scaling law MISO DSPM}ST under DSPM.]{\begin{centering}
\includegraphics[width=2.1in]{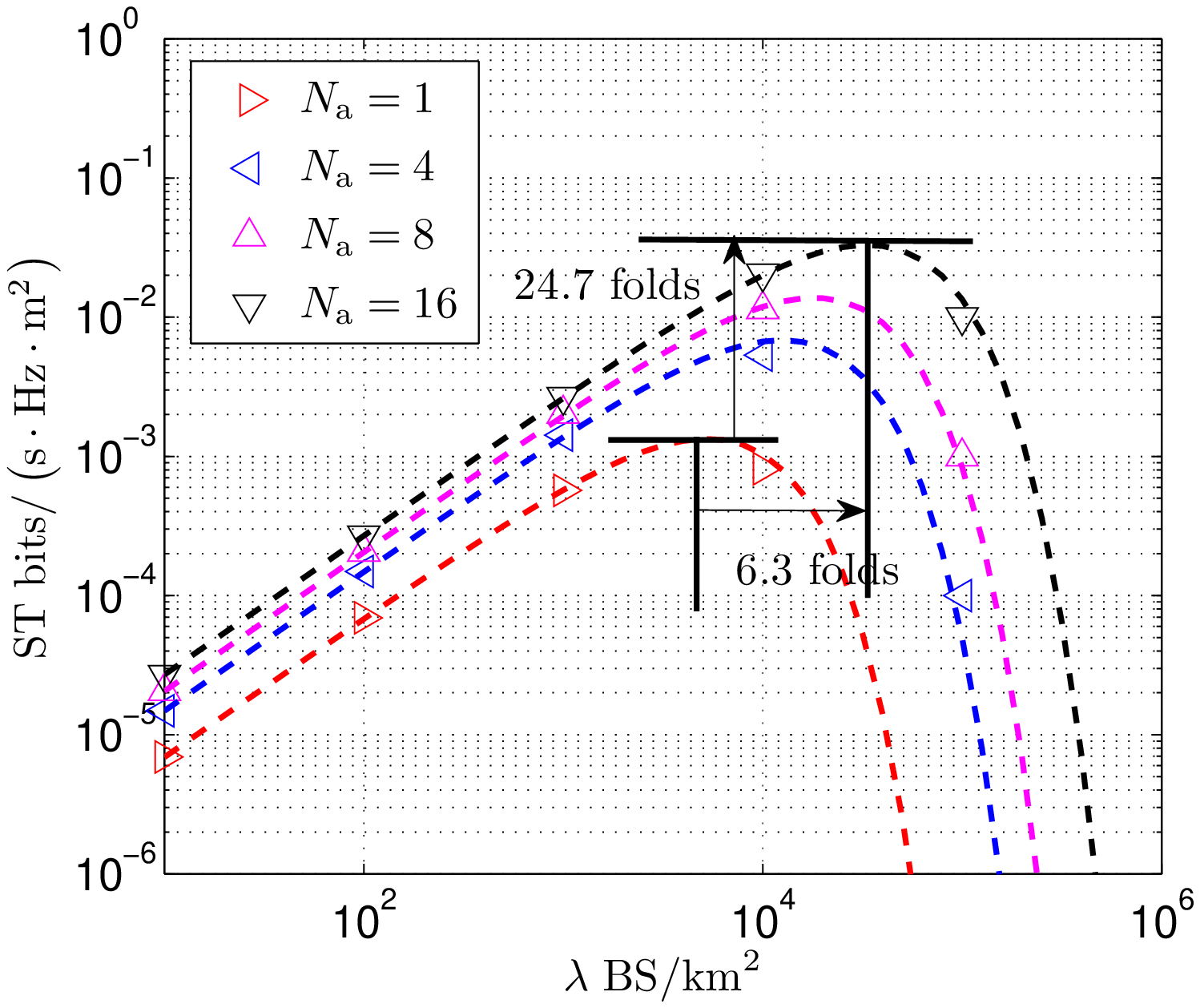}
\par\end{centering}
}
\par\end{centering}
\caption{\label{fig:CP and ST scaling law MISO}CP and ST varying with BS density
$\lambda$. For system settings, set $P=23$dBm, $\tau=10$dB and
$\Delta h$=2m. For SSPM, set $\alpha_{0}=4$. For DSPM, set $\alpha_{0}=2.5$,
$\alpha_{1}=4$ and $R_{1}=10$m.}
\end{figure*}

Theorem \ref{theorem: CP and ST scaling law MISO} indicates that
network densification would ultimately degrade both user and system
performance even when SU-BF is applied, which is somewhat pessimistic.
Nevertheless, adopting SU-BF at the BS side indeed significantly enhances
the desired signal power, which contributes to the improvement of
user performance. We use the results in Fig. \ref{fig:CP and ST scaling law MISO}
to illustrate this, which plots the CP and ST as a function of BS
density $\lambda$ under different $N_{\mathrm{a}}$. It is shown
from Fig. \ref{fig:CP scaling law MISO} that, besides increasing
CP, SU-BF could make CP degrade at a greater BS density under SSPM
and DSPM. Meanwhile, it can be seen from Figs. \ref{fig:ST scaling law MISO SSPM}
and \ref{fig:ST scaling law MISO DSPM} the maximal ST could be improved
by 32.4 and 24.7 folds under SSPM and DSPM, respectively, when 16
antennas are equipped, compared to the single-antenna case. More importantly,
the critical density could be considerably increased by SU-BF as well.
It means that the bottleneck of network densification could be partially
relieved with the application of SU-BF.

In addition, through the comparison of CP and ST scaling laws in Fig.
\ref{fig:CP and ST scaling law MISO}, it is also observed that the
improvement of system performance is at the cost of the degeneration
of user experience. For instance, when $N_{\mathrm{a}}=16$, network
ST grows with BS density at $1\times10^{4}$BS/$\mathrm{km}^{2}$
(see Figs. \ref{fig:ST scaling law MISO SSPM} and \ref{fig:ST scaling law MISO DSPM}),
under which CP already starts to diminish with $\lambda$ (see Fig.
\ref{fig:CP scaling law MISO}). Therefore, besides ensuring the system
performance, it is also critical to guarantee the user experience
when planning the deployment of small cell networks.

\subsection{Tradeoff between user and system performance}

To balance the tradeoff between user and system performance, a CP
requirement $\varepsilon$ is set to guarantee the QoS of downlink
users as
\begin{equation}
\mathsf{CP}\left(\lambda\right)=\mathbb{P}\left\{ \mathsf{SIR}_{\mathrm{U}_{0}}>\tau\right\} >\varepsilon.\label{eq: define CP with QoS Req}
\end{equation}
It should be noted that, although the CP and ST scaling law analysis
could be made using the lower and upper bounds in Theorem \ref{theorem: CP and ST scaling law MISO},
it is still intractable to analytically obtain the closed-form expression
of the critical density due to the complicated form of ST given in
Corollary \ref{corollary: CP and ST MISO}. As a substitution, we
derive a simple but accurate approximation of CP in the following
proposition.

\begin{proposition}

When SU-BF is applied by each multi-antenna BS, the ST in downlink
small cell networks under MSPM in (\ref{eq:MUPM}) is approximated
by $\tilde{\mathsf{ST}}_{N}^{\mathrm{M}}\left(\lambda\right)=\lambda\tilde{\mathsf{CP}}_{N}^{\mathrm{M}}\left(\lambda\right)\log_{2}\left(1+\tau\right)$,
where $\tilde{\mathsf{CP}}_{N}^{\mathrm{M}}\left(\lambda\right)$
is given by (\ref{eq:CP general MISO APP}).

\begin{equation}
\tilde{\mathsf{CP}}_{N}^{\mathrm{M}}\left(\lambda\right)=\left\{ \begin{array}{cc}
\frac{1}{1+\delta\left(\tau^{\dagger},\alpha_{0}\right)}\exp\left(-\pi\lambda\delta\left(\tau^{\dagger},\alpha_{0}\right)\Delta h^{2}\right), & N=1\\
\stackrel[n=0]{N-1}{\sum}\mathbb{E}_{r_{0}\in\left[R_{n},R_{n+1}\right)}\left\{ \exp\left[-\pi\lambda\left(\bar{R}_{n+1}^{2}\omega_{2}\left(\frac{\bar{R}_{n+1}^{\alpha_{n}}}{\tau^{\dagger}d_{0}^{\alpha_{n}}},\alpha_{n}\right)-d_{0}^{2}\omega_{2}\left(\frac{1}{\tau^{\dagger}},\alpha_{n}\right)\right.\right.\right.\\
+\left.\left.\left.\stackrel[i=n+1]{N-1}{\sum}\left(\bar{R}_{i+1}^{2}\omega_{2}\left(\frac{\bar{R}_{i+1}^{\alpha_{i}}}{\tau^{\dagger}K_{i}d_{0}^{\alpha_{n}}},\alpha_{i}\right)-\bar{R}_{i}^{2}\omega_{2}\left(\frac{\bar{R}_{i}^{\alpha_{i}}}{\tau^{\dagger}K_{i}d_{0}^{\alpha_{n}}},\alpha_{i}\right)\right)\right)\right]\right\} , & N>1
\end{array}\right.\label{eq:CP general MISO APP}
\end{equation}

In (\ref{eq:CP general MISO APP}), $\tau^{\dagger}=\frac{\tau}{N_{\mathrm{a}}}$,
$d_{0}=\sqrt{r_{0}^{2}+\Delta h^{2}}$ and the PDF of $r_{0}$ is
given by (\ref{eq:PDF of r0}).

\label{proposition: CP and ST MISO approximation}

\end{proposition}

\textit{Proof}: Please refer to Appendix \ref{subsec:Proof for MISO APP}.\qed

\begin{remark}

The approximation in Proposition \ref{proposition: CP and ST MISO approximation}
is used to provide a tractable approach to evaluate the performance
of dense small cell networks. The key is to use $\tilde{g}_{0}\sim\mathrm{Exp}\left(\frac{1}{2N_{\mathrm{a}}}\right)$
to approximate $\left\Vert \mathbf{h}_{\mathrm{U}_{0},\mathrm{BS}_{0}}\mathbf{v}_{\mathrm{U}_{0},\mathrm{BS}_{0}}^{\mathrm{T}}\right\Vert ^{2}\sim\chi_{2N_{\mathrm{a}}}^{2}$.
In particular, $\tilde{g}_{0}$ and $\left\Vert \mathbf{h}_{\mathrm{U}_{0},\mathrm{BS}_{0}}\mathbf{v}_{\mathrm{U}_{0},\mathrm{BS}_{0}}^{\mathrm{T}}\right\Vert ^{2}$
share the identical mean. Although higher moments of $\tilde{g}_{0}$
and $\left\Vert \mathbf{h}_{\mathrm{U}_{0},\mathrm{BS}_{0}}\mathbf{v}_{\mathrm{U}_{0},\mathrm{BS}_{0}}^{\mathrm{T}}\right\Vert ^{2}$
are close only when $N_{\mathrm{a}}$ is small, the impact of pathloss,
which is a dominating factor to channel gain, on CP and ST significantly
overwhelms that of small-scale fading when the distance from Tx's
to Rx's is small in UDN. For the above reason, the approximation is
reasonable.

\label{remark: CP and ST MISO approximation}

\end{remark}

\begin{figure*}[t]
\centering{}\subfloat[\label{fig:CP scaling law MISO - approximation}CP.]{\begin{centering}
\includegraphics[width=2.1in]{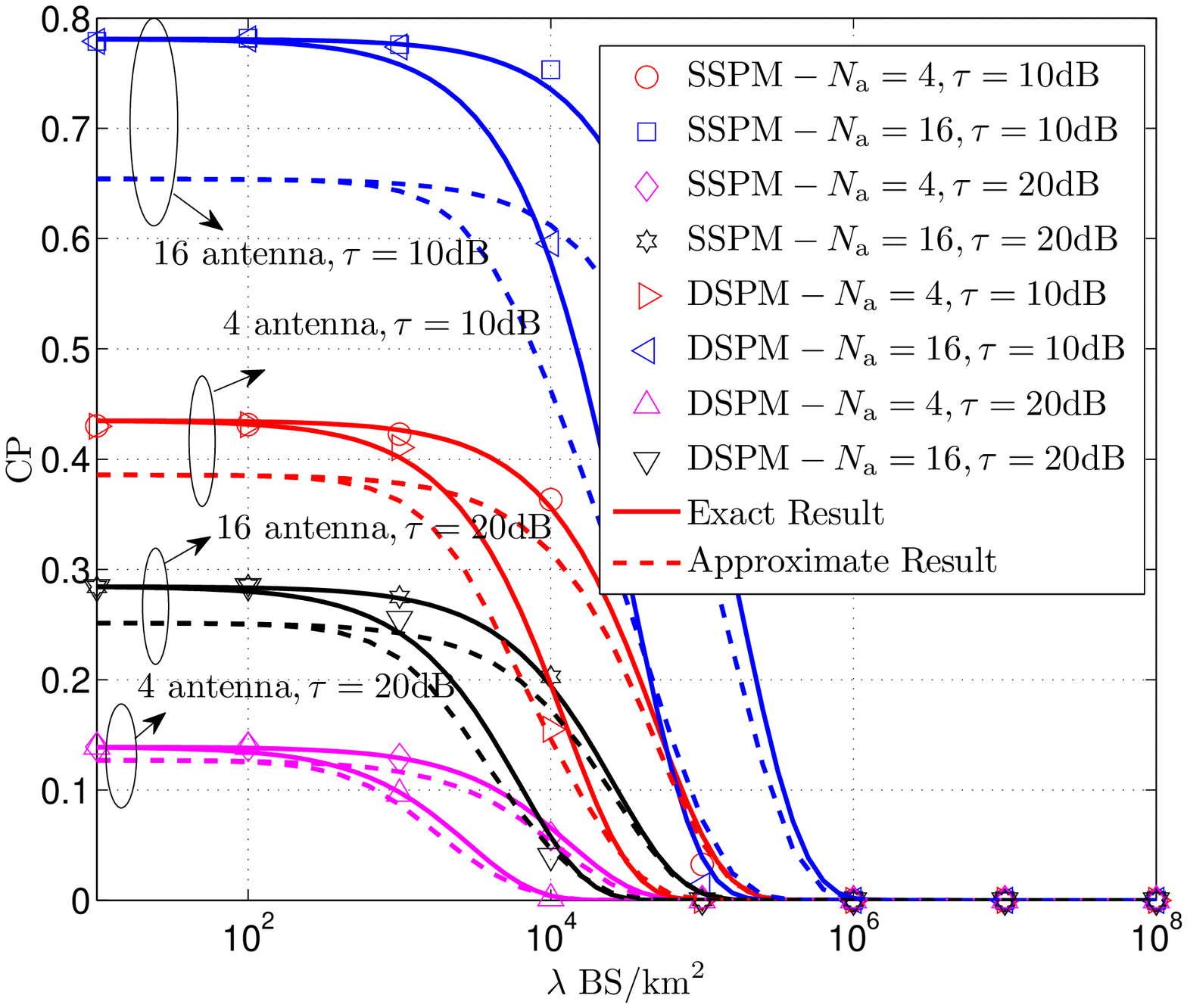}
\par\end{centering}
}\subfloat[\label{fig:ST scaling law MISO SSPM - approximation}ST under SSPM.]{\begin{centering}
\includegraphics[width=2.1in]{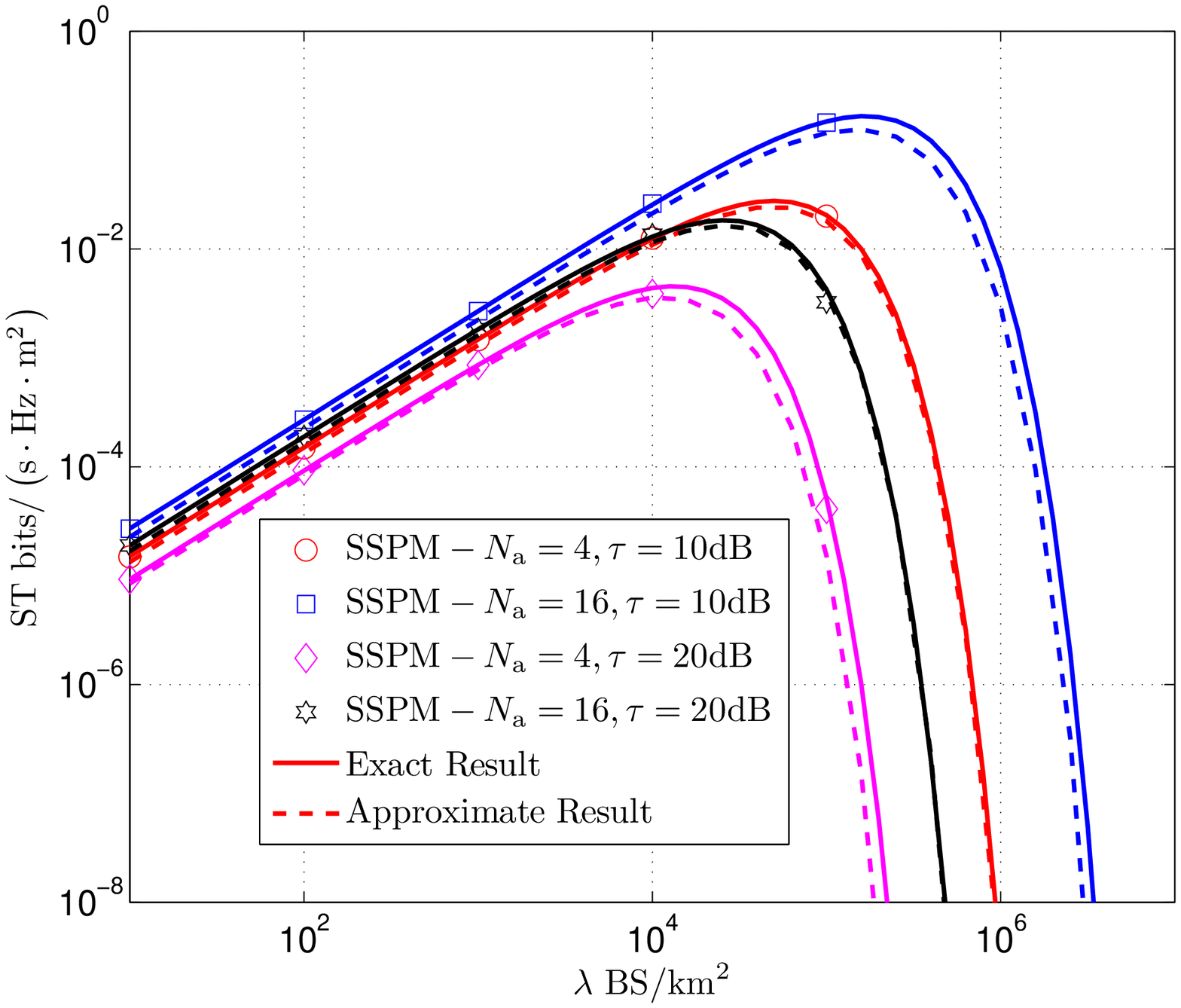}
\par\end{centering}
}\subfloat[\label{fig:ST scaling law MISO DSPM - approximation}ST under DSPM.]{\begin{centering}
\includegraphics[width=2.1in]{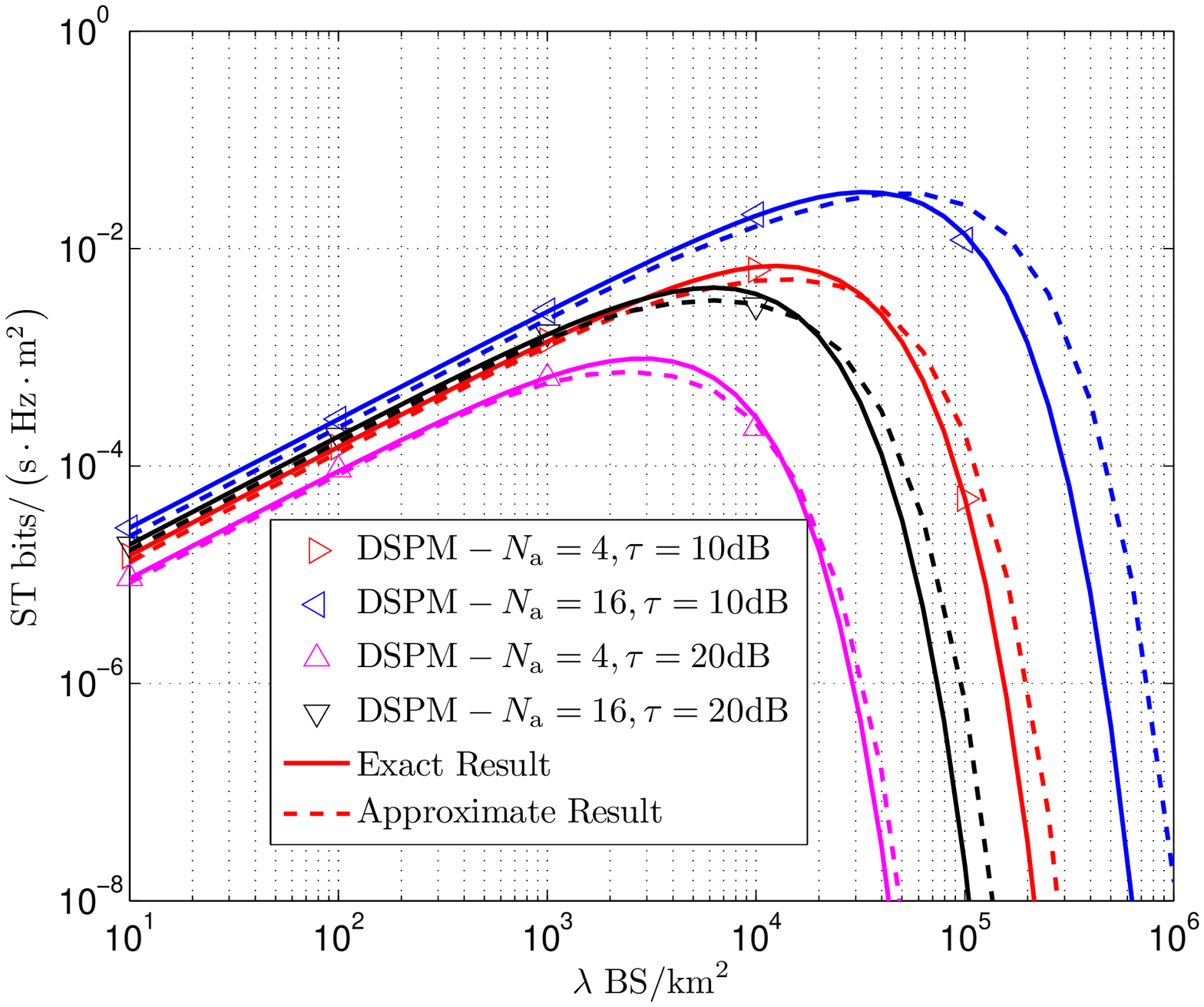}
\par\end{centering}
}\caption{\label{fig:CP and ST scaling law MISO - approximation}Exact and approximate
CP/ST varying with BS density $\lambda$. For system settings, For
system settings, set $P=23$dBm and $\Delta h$=2m. For SSPM, set
$\alpha_{0}=4$. For DSPM, set $\alpha_{0}=2.5$, $\alpha_{1}=4$
and $R_{1}=10$m.}
\end{figure*}

In Fig. \ref{fig:CP and ST scaling law MISO - approximation}, we
examine the accuracy of the approximations in Proposition \ref{proposition: CP and ST MISO approximation},
where the comparison between the exact and approximate results on
CP and ST is made. Fig. \ref{fig:CP scaling law MISO - approximation}
indicates that there exist gaps between the exact and approximate
results especially when BS density is small. Worsestill, the gaps
would be enlarged when the number of antennas equipped on each BS
is increased. However, the scaling behaviors of the approximate CP
and ST are identical to those of the exact results. Meanwhile, the
gaps are shown to be reduced when BS density becomes larger (see Figs.
\ref{fig:ST scaling law MISO SSPM - approximation} and \ref{fig:ST scaling law MISO DSPM - approximation}).
More importantly, the exact and approximate critical densities are
almost identical. As the approximation is used to derive more simple
and insightful results on the critical density, the above discussion
is sufficient to validate the accuracy of the approximation in Proposition
\ref{proposition: CP and ST MISO approximation}.

With the approximate results in Proposition \ref{proposition: CP and ST MISO approximation},
we then study the critical density with the CP requirement specified
in (\ref{eq: define CP with QoS Req}). From (\ref{eq: define CP with QoS Req}),
it is intuitive that whether or not the requirement could be satisfied
primarily depends on the deployment density of BSs. However, as observed
from Fig. \ref{fig:CP scaling law MISO - approximation}, even when
BS density is small, the maximal CP that can be achieved only reaches
0.79 under $N_{\mathrm{a}}=16$ and $\tau=10$dB. This indicates that,
besides BS density, other parameters such as pathloss exponents, decoding
threshold, etc., impact whether the CP requirement could be met as
well. In this light, we first analyze the necessary condition to acquire
the CP requirement. Afterward, we derive the critical density, under
which network ST can be maximized under the pre-set CP requirement.

\subsection{Critical density under SSPM and DSPM}

Aided by the approximation in Proposition \ref{proposition: CP and ST MISO approximation},
we first analyze the necessary condition, under which the CP requirement
specified in (\ref{eq: define CP with QoS Req}) could be satisfied.

\begin{theorem}

Under MSPM in (\ref{eq:MUPM}), the necessary condition to satisfy
the CP requirement in (\ref{eq: define CP with QoS Req}) is given
by
\begin{equation}
\frac{2\tau^{\dagger}\omega_{1}\left(\tau^{\dagger},\alpha_{N-1}\right)}{\alpha_{N-1}-2}<\varepsilon^{-1}-1,\label{eq: necessary condition MSPM}
\end{equation}
where $\tau^{\dagger}=\frac{\tau}{N_{\mathrm{a}}}$.

\label{theorem: feasible region MSPM}

\end{theorem}

\textit{Proof}: Please refer to Appendix \ref{subsec:Proof for Theorem feasible region MSPM}.\qed

Theorem \ref{theorem: feasible region MSPM} provides a direct approach
on how to reasonably adjust system parameters to meet the pre-set
CP requirement of downlink users. For instance, it is easy to prove
that $\psi_{N}\left(\tau,\alpha_{N-1},N_{\mathrm{a}}\right)=\frac{\tau}{N_{\mathrm{a}}}\omega_{1}\left(\frac{\tau}{N_{\mathrm{a}}},\alpha_{N-1}\right)$
in the left-hand-side of (\ref{eq: necessary condition MSPM}) is
an increasing function of the decoding threshold $\tau$. Therefore,
the CP requirement is less likely to be met with a greater $\tau$.
Nevertheless, Theorem \ref{theorem: feasible region MSPM} indicates
that increasing the number of antennas would directly relieve this.
Specifically, following (\ref{eq: necessary condition MSPM}), increasing
$N_{\mathrm{a}}$ is equivalent to lowering the decoding threshold
$\tau$. Although the results in Theorem \ref{theorem: feasible region MSPM}
are derived from the approximate results in Proposition \ref{proposition: CP and ST MISO approximation},
the benefits of applying multi-antenna techniques in UDN can be more
directly revealed.

Aided by Theorem \ref{theorem: feasible region MSPM}, we further
investigate the critical BS density under two typical pathloss models,
namely, SSPM in (\ref{eq:SUPM}) and DSPM in (\ref{eq:DUPM}), thereby
providing helpful insights and guideline towards the deployment of
dense small cell networks.

\begin{corollary}[Critical Density under SSPM]

Under SSPM in (\ref{eq:SUPM}), the critical BS density $\lambda_{1}^{\dagger}$,
under which network ST is maximized without the CP constraint, is
given by
\begin{equation}
\lambda_{1}^{\dagger}=\frac{\alpha_{0}-2}{2\pi\tau^{\dagger}\omega_{1}\left(\tau^{\dagger},\alpha_{0}\right)\Delta h^{2}},\label{eq:critical density wo Req SSPM}
\end{equation}
where $\tau^{\dagger}=\frac{\tau}{N_{\mathrm{a}}}$. With the CP constraint
$\varepsilon$, the critical BS density $\lambda_{1}^{*}$ is given
by
\begin{equation}
\lambda_{1}^{*}=\max\left(\lambda_{1}^{\dagger}\ln\left[\varepsilon^{-1}\left(1+\frac{2\tau^{\dagger}\omega_{1}\left(\tau^{\dagger},\alpha_{0}\right)}{\alpha_{0}-2}\right)^{-1}\right],\lambda_{1}^{\dagger}\right).\label{eq:critical density with Req SSPM}
\end{equation}

\label{corollary: critical density SSPM}

\end{corollary}

\textit{Proof}: Please refer to Appendix \ref{subsec:Proof for critical density SSPM}.\qed

Corollary \ref{corollary: critical density SSPM} reveals the fundamental
limitation of small cell networks by quantifying how many BSs could
be deployed per unit area (critical density), and more importantly
characterizes the impact of key system parameters on the critical
density. For instance, as $\psi\left(\tau,\alpha_{0},N_{\mathrm{a}}\right)=\frac{\tau}{N_{\mathrm{a}}}\omega_{1}\left(\frac{\tau}{N_{\mathrm{a}}},\alpha_{0}\right)$
in the denominator of $\lambda_{1}^{*}$ and $\lambda_{1}^{\dagger}$
decreases with the increasing $N_{\mathrm{a}}$, the results indicate
that increasing the number of antennas would result in an increase
of the critical density.

Next, we further investigate the critical densities under DSPM in
Corollary \ref{corollary: critical density DSPM}.

\begin{corollary}[Critical Density under DSPM]

Under DSPM in (\ref{eq:DUPM}), the critical BS density $\lambda_{2}^{\dagger}$,
under which network ST is maximized without the CP constraint, is
approximated as
\begin{equation}
\lambda_{2}^{\dagger}=\frac{1}{\pi\left[R_{1}^{2}\left(1+\delta\left(\tau^{\dagger},\alpha_{1}\right)\right)+\Delta h^{2}\delta\left(\tau^{\dagger},\alpha_{1}\right)\right]},\label{eq:critical density wo Req DSPM}
\end{equation}
where $\tau^{\dagger}=\frac{\tau}{N_{\mathrm{a}}}$. With the CP constraint
$\varepsilon$, the critical BS density $\lambda_{2}^{*}$ is approximated
as
\begin{equation}
\lambda_{2}^{*}=\max\left(\lambda_{2}^{\dagger}\ln\left[\varepsilon^{-1}\left(1+\delta\left(\tau^{\dagger},\alpha_{1}\right)\right)^{-1}\right],\lambda_{2}^{\dagger}\right).\label{eq:critical density with Req DSPM}
\end{equation}
The approximations are of high accuracy when $R_{1}$ in (\ref{eq:DUPM})
is large.

\label{corollary: critical density DSPM}

\end{corollary}

\textit{Proof}: Please refer to Appendix \ref{subsec:Proof for critical density DSPM}.
\qed

\begin{figure}[t]
\begin{centering}
\subfloat[\label{fig:critical density - test APP 1}ST under $R_{1}=10$m.]{\begin{centering}
\includegraphics[width=3in]{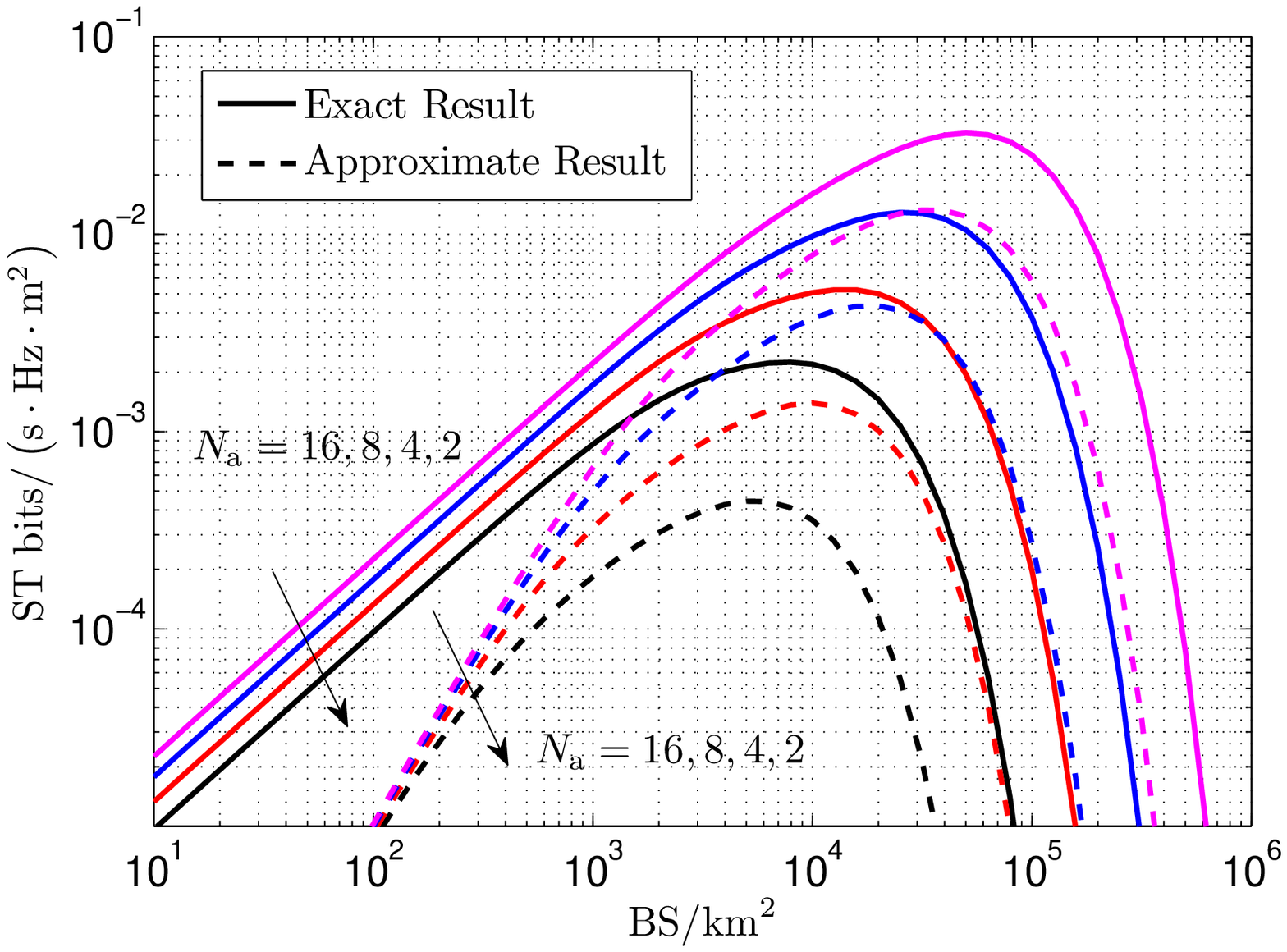}
\par\end{centering}
}\subfloat[\label{fig:critical density - test APP 2}ST under $R_{1}=50$m]{\begin{centering}
\includegraphics[width=3in]{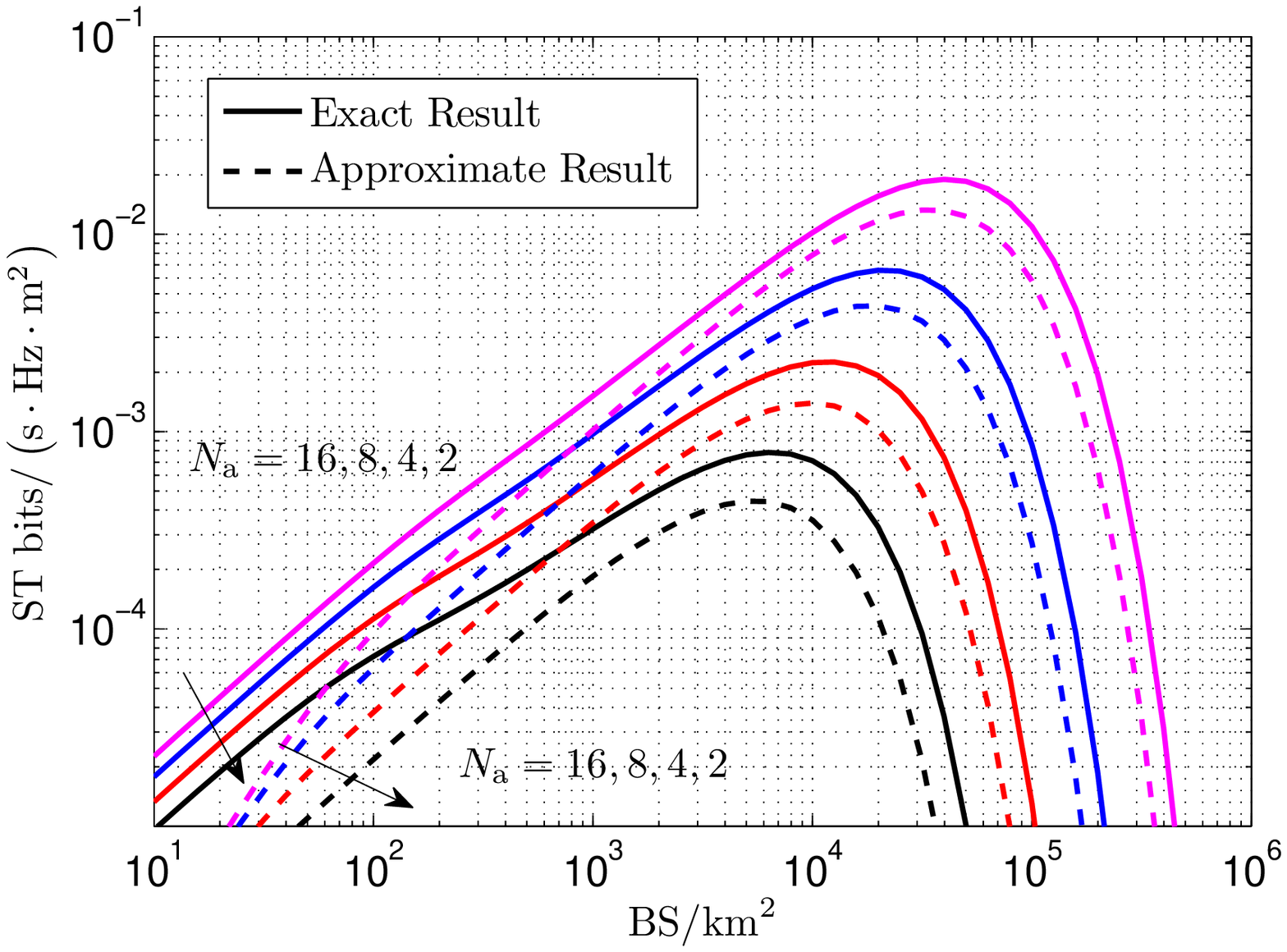}
\par\end{centering}
}
\par\end{centering}
\caption{\label{fig:critical density - test APP}Exact and approximate ST varying
with BS density $\lambda$ under DSPM. For system settings, set $P=23$dBm,
$\tau=0$dB and $\Delta h$=2m. For SSPM, set $\alpha_{0}=4$. For
DSPM, set $\alpha_{0}=2.5$ and $\alpha_{1}=4$.}
\end{figure}

As indicated by Corollary \ref{corollary: critical density DSPM},
the approximation of critical densities are valid when the corner
distance $R_{1}$ in (\ref{eq:DUPM}) is large. We use the results
in Fig. \ref{fig:critical density - test APP} to verify this. Figs.
\ref{fig:critical density - test APP 1} and \ref{fig:critical density - test APP 2}
plot the exact and approximate ST as a function of BS density under
DSPM when $R_{1}=10$m and $R_{1}=50$m, respectively. Notably, it
can be seen that the critical densities obtained via exact and approximate
results are almost identical under the given settings. According to
\cite{Ref_Two_Ray_Model}, $R_{1}\approx\frac{4h_{\mathrm{T}}h_{\mathrm{R}}f_{\mathrm{c}}}{c}$,
where $f_{c}$ denotes the carrier frequency and $c=3\times10^{8}$m/s
denotes the light speed. Given $h_{\mathrm{T}}=2.5$m and $h_{\mathrm{R}}=1.5$m,
$R_{1}$ basically ranges from several meters to dozens of meters
under sub-6GHz and increases with $f_{c}$. For this reason, the approximations
in Corollary \ref{corollary: critical density DSPM} are reasonable
in practice.

\begin{figure}[t]
\centering{}\subfloat[\label{fig:critical density SSPM}Critical density under SSPM.]{\begin{centering}
\includegraphics[width=3in]{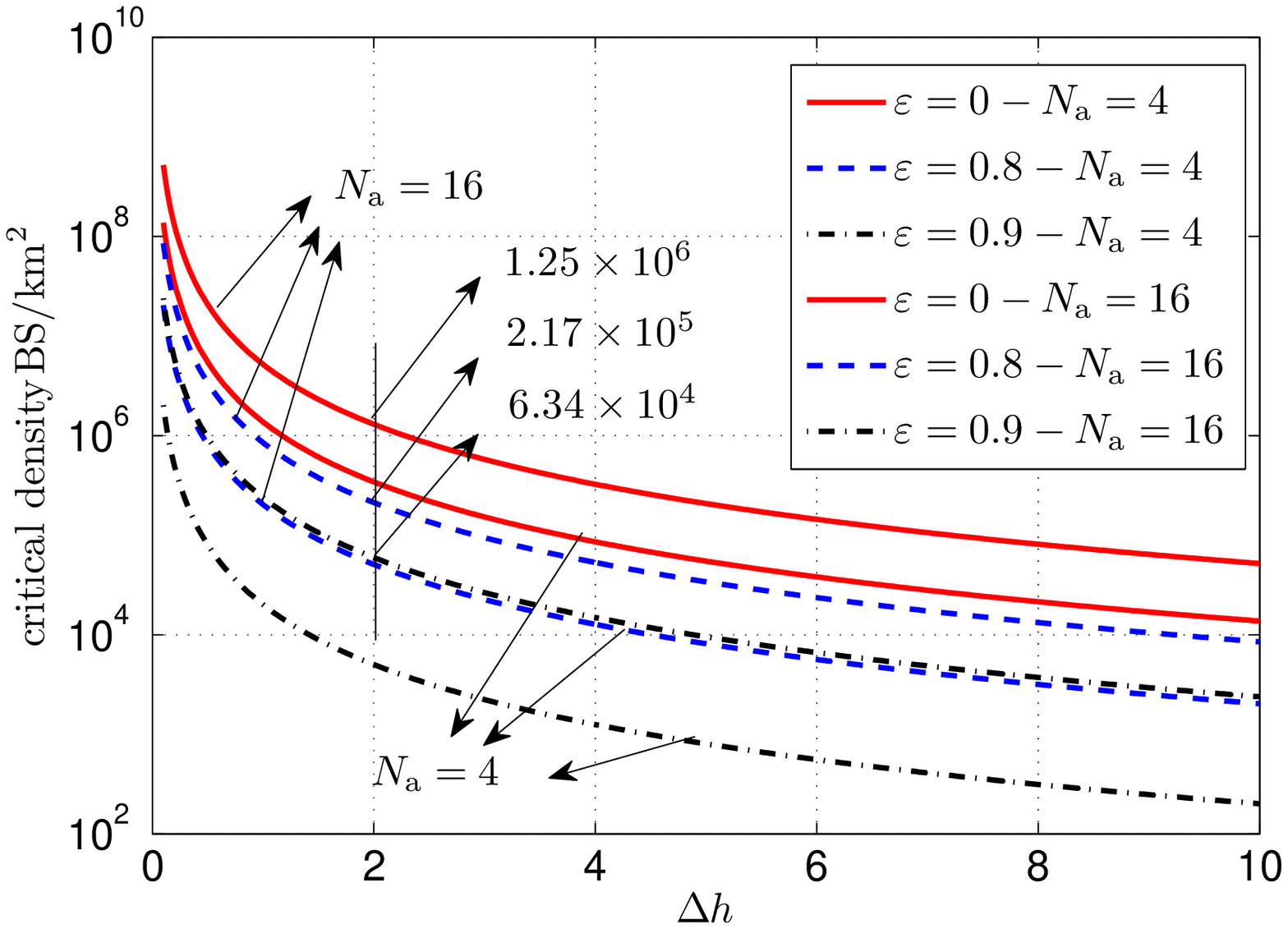}
\par\end{centering}
}\subfloat[\label{fig:critical density DSPM}Critical density under DSPM.]{\begin{centering}
\includegraphics[width=3in]{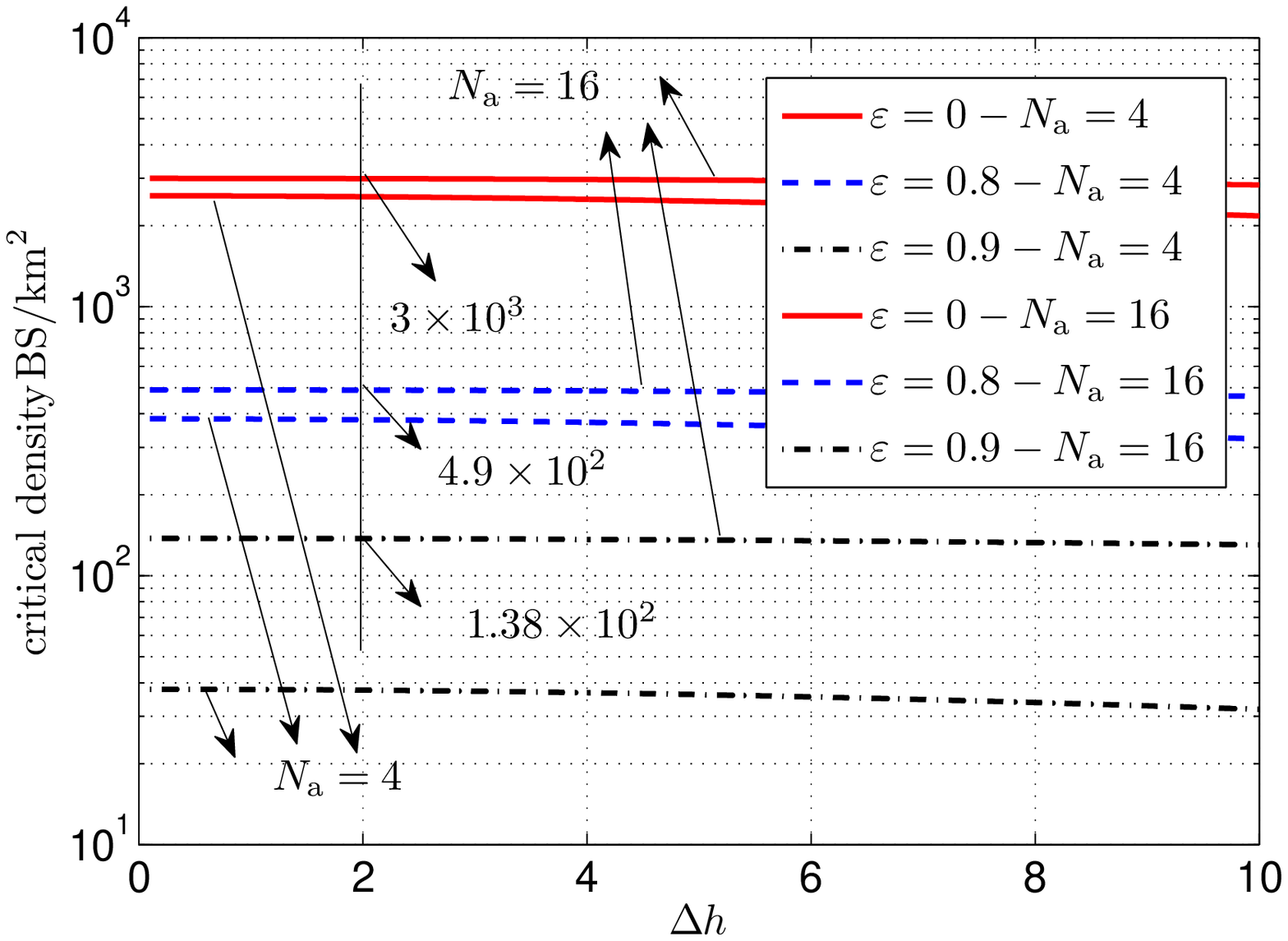}
\par\end{centering}
}\caption{\label{fig:critical density}Critical densities $\lambda^{*}$ and
$\lambda^{\dagger}$ varying with the AHD $\Delta h$. For system
settings, set $P=23$dBm and $\tau=0$dB. For SSPM, set $\alpha_{0}=4$.
For DSPM, set $\alpha_{0}=2.5$, $\alpha_{1}=4$ and $R_{1}=10$m.
Note that $\varepsilon=0$ is equivalent to the case, where no CP
requirement is considered.}
\end{figure}

It is also observed from Corollaries \ref{corollary: critical density SSPM}
and \ref{corollary: critical density DSPM} that the critical densities
would be decreased by the CP requirement $\varepsilon$ and AHD $\Delta h$.
Especially, Fig. \ref{fig:critical density} shows the critical density
as a function of $\Delta h$ under different $\varepsilon$. It is
observed from Fig. \ref{fig:critical density SSPM} that the critical
density is reduced by 5.8 and even 19.7 folds when $\varepsilon=0.8$
and $\varepsilon=0.9$, respectively, when $N_{\mathrm{a}}=16$ and
$\Delta h=2$m under SSPM. Using the same system settings, the critical
density is reduced by 6.1 and 21.7 folds, respectively, under DSPM
as well. The results demonstrate that the CP requirement greatly limits
the maximal BS deployment density. In addition, as critical density
would increase inversely with $\Delta h^{2}$, the above results also
reveal the essential influence of AHD on the BS deployment in downlink
small cell networks. In particular, it indicates that a great $\Delta h$
would hinder the increase of ST in dense scenarios. From this perspective,
it suggests that the antenna height of small cell BSs should be lowered,
thereby facilitating the maximization of network ST while ensuring
the QoS of downlink users.

\section{Conclusion\label{sec:Conclusion}}

In this paper, we have explored the fundamental limits of network
densification in downlink small cell networks when SU-BF serves as
the multi-antenna transmission technique under a generalized multi-slope
pathloss model. While incapable of improving the CP and ST scaling
laws, the application of MISO is shown to significantly enhance user
experience and system performance and even increase the critical density.
Meanwhile, aided by the simple but accurate approximations, the influence
of multi-antenna techniques on CP and ST could be explicitly revealed.
In addition, it is observed that the CP of downlink users starts to
diminish with the BS density when network ST is increased. Therefore,
to strike a better balance between user and system performance, we
have analyzed the critical density, under which network ST can be
maximized with the pre-set CP requirement. The results could provide
helpful guidance for the network deployment and application of network
densification in future wireless networks.

\appendix

\section{*}

\subsection{Proof for Proposition \ref{proposition: CP and ST SISO}\label{subsec:Proof for SP and ST SISO}}

In the following, $H_{\mathrm{U}_{0},\mathrm{BS}_{i}}$ is used to
replace $\left\Vert h_{\mathrm{U}_{0},\mathrm{BS}_{i}}\right\Vert ^{2}$.
Substituting (\ref{eq:SIR expression SISO}) into (\ref{eq: define CP}),
we have
\begin{equation}
\mathsf{CP}_{N}^{\mathrm{S}}\left(\lambda\right)\overset{\left(\mathrm{a}\right)}{=}\mathbb{P}\left\{ H_{\mathrm{U}_{0},\mathrm{BS}_{0}}>s_{N}^{\mathrm{S}}I_{\mathrm{IC}}^{\mathrm{S}}\right\} =\mathbb{E}_{d_{0}}\left[e^{-2\pi\lambda\int_{d_{0}}^{\infty}x\left(1-\frac{1}{1+s_{N}^{\mathrm{S}}Pl_{N}\left(x\right)}\right)\mathrm{d}x}\right],\label{eq:CP general proof SISO 1}
\end{equation}
where $s_{N}^{\mathrm{S}}=\frac{\tau}{Pl_{N}\left(d_{0}\right)}$.
For the derivation of step (a), please refer to (5) in \cite{Ref_SBPM}
for detail.

Given $N=1$, it is straightforward to obtain $s_{1}^{\mathrm{S}}=\frac{\tau d_{0}^{\alpha_{0}}}{P}$
and
\begin{align}
\mathsf{CP}_{1}^{\mathrm{S}}\left(\lambda\right)= & \mathbb{E}_{d_{0}}\left[\exp\left(-\frac{2\pi\lambda\tau\omega_{1}\left(\tau,\alpha_{0}\right)}{\alpha-2}d_{0}^{2}\right)\right]=\mathbb{E}_{r_{0}}\left[\exp\left(-\frac{2\pi\lambda\tau\omega_{1}\left(\tau,\alpha_{0}\right)}{\alpha-2}\left(r_{0}^{2}+\Delta h^{2}\right)\right)\right]\nonumber \\
\overset{\left(\mathrm{a}\right)}{=} & \frac{1}{1+\delta\left(\tau,\alpha_{0}\right)}\exp\left(-\pi\lambda\bigtriangleup h^{2}\delta\left(\tau,\alpha_{0}\right)\right),\label{eq:CP general proof SISO 2}
\end{align}
where (a) follows because the PDF of $r_{0}$ is given by (\ref{eq:PDF of r0}).

Given $N>1$ and $d_{0}\in\left[\bar{R}_{n},\bar{R}_{n+1}\right)$
with $\bar{R}_{n}=\sqrt{r_{0}^{2}+R_{n}^{2}}$, $\int_{d_{0}}^{\infty}x^{k-1}\left(1-\frac{1}{1+s_{N}^{\mathrm{S}}Pl_{N}\left(x\right)}\right)\mathrm{d}x$
in (\ref{eq:CP general proof SISO 1}) turns into $\int_{d_{0}}^{\infty}x\left(1-\frac{1}{1+s_{N}^{\mathrm{S}}Pl_{N}\left(x\right)}\right)\mathrm{d}x$\begin{small}
\begin{align}
= & \int_{d_{0}}^{\bar{R}_{n+1}}x\left(1-\frac{1}{1+\tau d_{0}^{\alpha_{n}}x^{-\alpha_{n}}}\right)\mathrm{d}x+\stackrel[i=n+1]{N-1}{\sum}\int_{\bar{R}_{i}}^{\bar{R}_{i+1}}x\left(1-\frac{1}{1+\tau K_{i}d_{0}^{\alpha_{n}}x^{-\alpha_{i}}}\right)\mathrm{d}x\nonumber \\
= & \frac{1}{2}\left[\bar{R}_{n+1}^{2}\omega_{2}\left(\frac{\bar{R}_{n+1}^{\alpha_{n}}}{\tau d_{0}^{\alpha_{n}}},\alpha_{n}\right)-d_{0}^{2}\omega_{2}\left(\tau^{-1},\alpha_{n}\right)\right]+\stackrel[i=n+1]{N-1}{\sum}\left[\frac{\bar{R}_{i+1}^{2}}{2}\omega_{2}\left(\frac{\bar{R}_{i+1}^{\alpha_{i}}}{\tau K_{i}d_{0}^{\alpha_{n}}},\alpha_{i}\right)-\frac{\bar{R}_{i}^{2}}{2}\omega_{2}\left(\frac{\bar{R}_{i}^{\alpha_{i}}}{\tau K_{i}d_{0}^{\alpha_{n}}},\alpha_{i}\right)\right].\label{eq:CP general proof SISO 3}
\end{align}
\end{small}According to (\ref{eq:CP general proof SISO 3}), $\mathsf{CP}_{N}^{\mathrm{S}}\left(\lambda\right)$
$\left(N>1\right)$ could be obtained and hence the proof is completed.

\subsection{Proof for Theorem \ref{theorem: CP and ST scaling law SISO}\label{subsec:Proof for scaling law}}

From (\ref{eq:CP general SISO}) in Proposition \ref{proposition: CP and ST SISO},
the proof for the scaling laws of CP and ST under the SSPM is straightforward.
Therefore, we focus on the proof for the case with $N>1$.

Given $N>1$, the CP in (\ref{eq:CP general SISO}) can be expressed
as $\mathsf{CP}_{N}^{\mathrm{S}}\left(\lambda\right)$
\begin{equation}
=\mathbb{E}_{r_{0}\in\left[R_{0},R_{N-1}\right)}\left[e^{-2\pi\lambda\int_{d_{0}}^{\infty}x\left(1-\frac{1}{1+s_{N}^{\mathrm{S}}Pl_{N}\left(x\right)}\right)\mathrm{d}x}\right]+\mathbb{E}_{r_{0}\in\left[R_{N-1},R_{N}\right)}\left[e^{-2\pi\lambda\int_{d_{0}}^{\infty}x\left(1-\frac{1}{1+s_{N}^{\mathrm{S}}Pl_{N}\left(x\right)}\right)\mathrm{d}x}\right].\label{eq:bound proof SISO 1}
\end{equation}

Then, the following inequality holds true, i.e.,
\begin{align}
\mathsf{CP}_{N}^{\mathrm{S}}\left(\lambda\right)> & \mathbb{E}_{r_{0}\in\left[R_{N-1},R_{N}\right)}\left[e^{-2\pi\lambda\int_{d_{0}}^{\infty}x\left(1-\frac{1}{1+s_{N}^{\mathrm{S}}Pl_{N}\left(x\right)}\right)\mathrm{d}x}\right].\label{eq:bound proof SISO 2}
\end{align}
As $d_{0}=\sqrt{r_{0}^{2}+\Delta h^{2}}$, $\bar{R}_{N-1}=\sqrt{R_{N-1}^{2}+\Delta h^{2}}$
and $R_{N}=\infty$, when $d_{0}\in\left[\bar{R}_{N-1},\infty\right)$,
$s_{N}^{\mathrm{S}}=\frac{\tau}{PK_{N-1}d_{0}^{-\alpha_{N-1}}}$ and
$l_{N}\left(x\right)=K_{N-1}x^{-\alpha_{N-1}}$, the integral in (\ref{eq:bound proof SISO 2})
turns into
\begin{align}
 & \int_{d_{0}}^{\infty}x\left(1-\frac{1}{1+\tau d_{0}^{\alpha_{N-1}}x^{-\alpha_{N-1}}}\right)\mathrm{d}x=\frac{\delta\left(\tau,\alpha_{N-1}\right)d_{0}^{2}}{2}=\frac{\delta\left(\tau,\alpha_{N-1}\right)}{2}\left(r_{0}^{2}+\Delta h^{2}\right),\label{eq:bound proof SISO 3}
\end{align}
where $\delta\left(\tau,\alpha_{N-1}\right)=\frac{2\tau\omega_{1}\left(\tau,\alpha_{N-1}\right)}{\alpha_{N-1}-2}$.
Following (\ref{eq:bound proof SISO 3}), we derive the lower bound
of $\mathsf{CP}_{N}^{\mathrm{S}}\left(\lambda\right)$ as
\begin{align}
\mathsf{CP}_{N}^{\mathrm{S}}\left(\lambda\right)> & \mathsf{CP}_{N-\mathrm{L}}^{\mathrm{\mathrm{S}}}\left(\lambda\right)=\mathbb{E}_{r_{0}\in\left[R_{N-1},\infty\right)}\left[e^{-\pi\lambda\delta\left(\tau,\alpha_{N-1}\right)\left(r_{0}^{2}+\Delta h^{2}\right)}\right]\nonumber \\
= & \frac{e^{-\pi\lambda\left[R_{N-1}^{2}+\delta\left(\tau,\alpha_{N-1}\right)\left(R_{N-1}^{2}+\Delta h^{2}\right)\right]}}{1+\delta\left(\tau,\alpha_{N-1}\right)}.\label{eq:bound proof SISO 3-1}
\end{align}
Therefore, it can be shown that $\exists\frac{1}{1+\delta\left(\tau,\alpha_{N-1}\right)}>0$,
$\forall\lambda>0$,
\begin{align}
\left|\mathsf{CP}_{N-\mathrm{L}}^{\mathrm{\mathrm{S}}}\left(\lambda\right)\right| & \geq\frac{e^{-\pi\lambda\left[R_{N-1}^{2}+\delta\left(\tau,\alpha_{N-1}\right)\left(R_{N-1}^{2}+\Delta h^{2}\right)\right]}}{1+\delta\left(\tau,\alpha_{N-1}\right)}.\label{eq:bound proof SISO 4}
\end{align}
According to Definition \ref{definition: scaling behavior}, $\mathsf{CP}_{N-\mathrm{L}}^{\mathrm{\mathrm{S}}}=\Omega\left(e^{-\pi\lambda\left[R_{N-1}^{2}+\delta\left(\tau,\alpha_{N-1}\right)\left(R_{N-1}^{2}+\Delta h^{2}\right)\right]}\right)$
holds true.

In the following, we analyze the upper bound of $\mathsf{CP}_{N}^{\mathrm{S}}\left(\lambda\right)$.
When $r_{0}\in\left[R_{n},R_{n+1}\right)$ or equivalently $d_{0}\in\left[\bar{R}_{n},\bar{R}_{n+1}\right)$
$\left(n=0,1,\ldots,N-2\right)$, $s_{N}^{\mathrm{S}}=\frac{\tau d_{0}^{\alpha_{n}}}{PK_{n}}$.
As such, $\int_{d_{0}}^{\infty}x\left(1-\frac{1}{1+s_{N}^{\mathrm{S}}Pl_{N}\left(x\right)}\right)\mathrm{d}x$
in the first term of (\ref{eq:bound proof SISO 1}) can be manipulated
as $\int_{d_{0}}^{\infty}x\left(1-\frac{1}{1+s_{N}^{\mathrm{S}}Pl_{N}\left(x\right)}\right)\mathrm{d}x$
\begin{align}
\overset{\left(\mathrm{a}\right)}{>} & \int_{\bar{R}_{N-1}}^{\infty}x\left(1-\frac{1}{1+\frac{\tau K_{N-1}}{K_{n}d_{0}^{-\alpha_{n}}}x^{-\alpha_{N-1}}}\right)\mathrm{d}x=\frac{\tau K_{N-1}\bar{R}_{N-1}^{2-\alpha_{N-1}}d_{0}^{\alpha_{n}}}{K_{n}\left(\alpha_{N-1}-2\right)}\omega_{1}\left(\frac{\tau K_{N-1}d_{0}^{\alpha_{n}}}{K_{n}R_{N-1}^{\alpha_{N-1}}},\alpha_{N-1}\right)\nonumber \\
\overset{\left(\mathrm{b}\right)}{>} & \frac{\tau K_{N-1}\bar{R}_{N-1}^{2-\alpha_{N-1}}\Delta h^{\alpha_{n}}}{K_{n}\left(\alpha_{N-1}-2\right)}\omega_{1}\left(\frac{\tau K_{N-1}}{K_{n}},\alpha_{N-1}\right)=q_{1}\left(n\right),\label{eq:bound proof SISO 4 - 1}
\end{align}
where (a) follows due to $d_{0}<\bar{R}_{N-1}$, and (b) follows because
$d_{0}>\Delta h$, $d_{0}^{\alpha_{n}}<R_{N-1}^{\alpha_{N-1}}$ and
$\omega_{1}\left(x,\alpha_{N-1}\right)$ is a decreasing function
of $x$ (see Lemma \ref{lemma: hypergeometric function}). Using (\ref{eq:bound proof SISO 4 - 1})
and the PDF of $r_{0}$ in (\ref{eq:PDF of r0}), we have $\mathbb{E}_{r_{0}\in\left[R_{0},R_{N-1}\right)}\left[e^{-2\pi\lambda\int_{d_{0}}^{\infty}x\left(1-\frac{1}{1+s_{N}^{\mathrm{S}}Pl_{N}\left(x\right)}\right)\mathrm{d}x}\right]$
\begin{align}
< & \stackrel[n=0]{N-2}{\sum}\mathbb{E}_{r_{0}\in\left[R_{n},R_{n+1}\right)}\left[e^{-2\pi\lambda q_{1}\left(n\right)}\right]=\stackrel[n=0]{N-2}{\sum}e^{-2\pi\lambda q_{1}\left(n\right)}\left(e^{-\pi\lambda R_{n}^{2}}-e^{-\pi\lambda R_{n+1}^{2}}\right).\label{eq:bound proof SISO 5}
\end{align}

When $r_{0}\in\left[R_{N-1},\infty\right)$, the second term of (\ref{eq:bound proof SISO 1})
is already given by $\mathsf{CP}_{N-\mathrm{L}}^{\mathrm{\mathrm{S}}}\left(\lambda\right)$
in (\ref{eq:bound proof SISO 3-1}). Hence, it is easy to obtain that
\begin{align}
\mathsf{CP}_{N}^{\mathrm{\mathrm{S}}}\left(\lambda\right)< & \stackrel[n=0]{N-2}{\sum}e^{-2\pi\lambda q_{1}\left(n\right)}\left(e^{-\pi\lambda R_{n}^{2}}-e^{-\pi\lambda R_{n+1}^{2}}\right)+\mathsf{CP}_{N-\mathrm{L}}^{\mathrm{\mathrm{S}}}\left(\lambda\right)\nonumber \\
< & \stackrel[n=0]{N-2}{\sum}e^{-2\pi\lambda q_{1}\left(n\right)}e^{-\pi\lambda R_{n}^{2}}+\mathsf{CP}_{N-\mathrm{L}}^{\mathrm{\mathrm{S}}}\left(\lambda\right)\overset{\left(\mathrm{a}\right)}{<}\stackrel[n=0]{N-2}{\sum}e^{-2\pi\lambda q_{1}\left(n\right)}+e^{-\pi\lambda R_{N-1}^{2}}\nonumber \\
= & \mathsf{CP}_{N-\mathrm{U}}^{\mathrm{\mathrm{S}}}\left(\lambda\right).\label{eq:bound proof SISO 6}
\end{align}
where (a) follows because $e^{-\pi\lambda R_{n}^{2}}<1$ and it is
direct to show $\mathsf{CP}_{N-\mathrm{L}}^{\mathrm{\mathrm{S}}}\left(\lambda\right)<e^{-\pi\lambda R_{N-1}^{2}}$.
In (\ref{eq:bound proof SISO 6}), if $n\in\mathbb{C}$ $\left(\mathbb{C}=\left\{ 0,1,\ldots,N-2\right\} \right)$,
which enables $2q_{1}\left(n\right)>R_{N-1}^{2}$, then the inequality
$e^{-2\pi\lambda q_{1}\left(n\right)}<e^{-\pi\lambda R_{N-1}^{2}}$
holds. Thus, $\mathsf{CP}_{N-\mathrm{U}}^{\mathrm{\mathrm{S}}}\left(\lambda\right)$
in (\ref{eq:bound proof SISO 6}) turns into
\begin{align*}
\mathsf{CP}_{N-\mathrm{U}}^{\mathrm{\mathrm{S}}}\left(\lambda\right)=\stackrel[n=0]{N-2}{\sum}e^{-2\pi\lambda q_{1}\left(n\right)}+e^{-\pi\lambda R_{N-1}^{2}}< & Ne^{-\pi\lambda R_{N-1}^{2}},
\end{align*}
which indicates that $\exists N>0$, $\forall\lambda>0$,
\begin{align}
\left|\mathsf{CP}_{N-\mathrm{U}}^{\mathrm{\mathrm{S}}}\left(\lambda\right)\right|< & Ne^{-\pi\lambda R_{N-1}^{2}}.\label{eq:bound proof SISO 7}
\end{align}

If $n\in\mathbb{C}^{\dagger}$ $\left(\mathbb{C}\subseteq\left\{ 0,1,\ldots,N-2\right\} \right)$,
which enables $2q_{1}\left(n\right)\leq R_{N-1}^{2}$, then we denote
$n=N^{\dagger}$, which makes $e^{-2\pi\lambda q_{1}\left(N^{\dagger}\right)}\geq e^{-2\pi\lambda q_{1}\left(n\right)}$
$\left(0\leq n\leq N-2\right)$. It is apparent that $e^{-2\pi\lambda q_{1}\left(N^{\dagger}\right)}\geq e^{-\pi\lambda R_{N-1}^{2}}$
holds as well. Thus, we have
\begin{align*}
\mathsf{CP}_{N-\mathrm{U}}^{\mathrm{\mathrm{S}}}\left(\lambda\right)=\stackrel[n=0]{N-2}{\sum}e^{-2\pi\lambda q_{1}\left(n\right)}+e^{-\pi\lambda R_{N-1}^{2}}< & Ne^{-2\pi\lambda q_{1}\left(N^{\dagger}\right)}.
\end{align*}
In this case, $\exists N>0$, $\forall\lambda>0$,
\begin{align}
\left|\mathsf{CP}_{N-\mathrm{U}}^{\mathrm{S}}\left(\lambda\right)\right| & <Ne^{-2\pi\lambda q_{1}\left(N^{\dagger}\right)}.\label{eq:bound proof SISO 8}
\end{align}
Following Definition \ref{definition: scaling behavior} and the results
in (\ref{eq:bound proof SISO 7}) and (\ref{eq:bound proof SISO 8}),
it can be shown that $\mathsf{CP}_{N-\mathrm{U}}^{\mathrm{\mathrm{S}}}\left(\lambda\right)=\mathcal{O}\left(e^{-\pi\lambda R_{N-1}^{2}}\right)$
or $\mathsf{CP}_{N-\mathrm{U}}^{\mathrm{\mathrm{S}}}\left(\lambda\right)=\mathcal{O}\left(e^{-2\pi\lambda q_{1}\left(N^{\dagger}\right)}\right)$
holds true.

According to the above scaling law analysis of $\mathsf{CP}_{N-\mathrm{U}}^{\mathrm{\mathrm{S}}}\left(\lambda\right)$
and $\mathsf{CP}_{N-\mathrm{L}}^{\mathrm{\mathrm{S}}}\left(\lambda\right)$,
it is easy to show that there exists a constant $\kappa$, which makes
$\mathsf{CP}_{N}^{\mathrm{S}}\left(\lambda\right)$ scale with $\lambda$
as $e^{-\kappa\lambda}$. Therefore, based on the definition of ST
in (\ref{eq: define ST}), $\mathsf{ST}_{N}^{\mathrm{S}}\left(\lambda\right)$
scales with $\lambda$ as $\lambda e^{-\kappa\lambda}$.

\subsection{Proof for Corollary \ref{corollary: CP and ST MISO}\label{subsec:Proof for Corollary CP and ST MISO}}

From (\ref{eq:SIR expression}) and (\ref{eq: define CP}), when SU-BF
is applied, the coverage probability $\mathsf{CP}_{N}^{\mathrm{M}}\left(\lambda\right)$
is given as
\begin{align}
\mathsf{CP}_{N}^{\mathrm{M}}\left(\lambda\right)= & \mathbb{P}\left\{ \left\Vert \mathbf{h}_{\mathrm{U}_{0},\mathrm{BS}_{0}}\mathbf{v}_{\mathrm{U}_{0},\mathrm{BS}_{0}}^{\mathrm{T}}\right\Vert ^{2}>s_{N}^{\mathrm{S}}I_{\mathrm{IC}}\right\} ,\label{eq:MISO proof 1}
\end{align}
where $s_{N}^{\mathrm{S}}=\frac{\tau}{Pl_{N}\left(d_{0}\right)}$.
As discussed in Section \ref{subsec:SU-BF}, $\left\Vert \mathbf{h}_{\mathrm{U}_{0},\mathrm{BS}_{0}}\mathbf{v}_{\mathrm{U}_{0},\mathrm{BS}_{0}}^{\mathrm{T}}\right\Vert \sim\chi_{2N_{\mathrm{a}}}^{2}$
and $\left\Vert \mathbf{h}_{\mathrm{U}_{0},\mathrm{BS}_{i}}\mathbf{v}_{\mathrm{U}_{0},\mathrm{BS}_{i}}^{\mathrm{T}}\right\Vert \sim\chi_{2}^{2}$
$\left(i\neq0\right)$. In consequence, we have
\begin{align}
\mathsf{CP}_{N}^{\mathrm{M}}\left(\lambda\right)\stackrel{\left(\mathrm{a}\right)}{=} & \mathbb{E}\left[\int_{0}^{\infty}\stackrel[k=0]{N_{\mathrm{a}}-1}{\sum}\frac{\left(xs\right)^{k}}{k!}e^{-xs}\mathrm{d}\mathbb{P}\left(I_{\mathrm{IC}}\leq x\right)\right]=\mathbb{E}\left[\stackrel[k=0]{N_{\mathrm{a}}-1}{\sum}\frac{\left(-s\right)^{k}}{k!}\frac{\mathrm{d}^{k}}{\mathrm{d}s^{k}}\mathcal{L}_{I_{\mathrm{IC}}}\left(s\right)\right],\label{eq:MISO proof 2}
\end{align}
where $s=\frac{s_{N}^{\mathrm{S}}}{2}$ and (a) follows by conditioning
on $I_{\mathrm{IC}}$ and calculating the complementary cumulative
distribution function of $\left\Vert \mathbf{h}_{\mathrm{U}_{0},\mathrm{BS}_{0}}\mathbf{v}_{\mathrm{U}_{0},\mathrm{BS}_{0}}^{\mathrm{T}}\right\Vert $.
$\mathcal{L}_{I_{\mathrm{IC}}}\left(s\right)$ denotes the Laplace
Transform of $I_{\mathrm{IC}}$ evaluated at $s$, which is given
by
\begin{align*}
\mathcal{L}_{I_{\mathrm{IC}}}\left(s\right)= & \mathbb{E}\left[\exp\left(-\underset{\tiny{\mathrm{BS}_{i}\in\tilde{\Pi}_{\mathrm{BS}}}}{\sum}sP\left\Vert \mathbf{h}_{\mathrm{U}_{0},\mathrm{BS}_{i}}\mathbf{v}_{\mathrm{U}_{0},\mathrm{BS}_{i}}^{\mathrm{T}}\right\Vert ^{2}l_{N}\left(d_{i}\right)\right)\right]=\underset{\tiny{\mathrm{BS}_{i}\in\tilde{\Pi}_{\mathrm{BS}}}}{\sum}\frac{1}{1+sPl_{N}\left(d_{i}\right)}\\
= & \exp\left(-2\pi\lambda\int_{d_{0}}^{\infty}x\left(1-\frac{1}{1+2sPl_{N}\left(x\right)}\right)\mathrm{d}x\right).
\end{align*}
Hence, the proof is complete.

\subsection{Proof for Theorem \ref{theorem: CP and ST scaling law MISO}\label{subsec:Proof for scaling law MISO}}

When SU-BF is applied, we first analyze the CP lower bound in the
following. According to the assumption that $\left\Vert \mathbf{h}_{\mathrm{U}_{0},\mathrm{BS}_{0}}\mathbf{v}_{\mathrm{U}_{0},\mathrm{BS}_{0}}^{\mathrm{T}}\right\Vert ^{2}\sim\chi_{2N_{\mathrm{a}}}^{2}$,
it is easy to show that
\begin{align}
\mathsf{CP}_{N}^{\mathrm{M}}\left(\lambda\right)= & \mathbb{P}\left\{ \left\Vert \mathbf{h}_{\mathrm{U}_{0},\mathrm{BS}_{0}}\mathbf{v}_{\mathrm{U}_{0},\mathrm{BS}_{0}}^{\mathrm{T}}\right\Vert ^{2}>s_{N}^{\mathrm{S}}I_{\mathrm{IC}}\right\} \overset{\left(\mathrm{a}\right)}{\geq}\mathbb{P}\left\{ g_{0}>s_{N}^{\mathrm{S}}I_{\mathrm{IC}}\right\} \nonumber \\
= & \mathsf{CP}_{N-\mathrm{L}}^{\mathrm{M}}\left(\lambda\right)=\mathsf{CP}_{N}^{\mathrm{S}}\left(\lambda\right),\label{eq:scaling MISO proof 1}
\end{align}
where $s_{N}^{\mathrm{S}}=\frac{\tau}{Pl_{N}\left(d_{0}\right)}$,
$g_{0}\sim\chi_{2}^{2}$ and $\mathsf{CP}_{N}^{\mathrm{S}}\left(\lambda\right)$
is given by Proposition \ref{proposition: CP and ST SISO}. The inequality
in (a) holds, since the degree of freedom of the chi-square distributed
random variable $\left\Vert \mathbf{h}_{\mathrm{U}_{0},\mathrm{BS}_{0}}\mathbf{v}_{\mathrm{U}_{0},\mathrm{BS}_{0}}^{\mathrm{T}}\right\Vert ^{2}$
is greater than or equal to that of $g_{0}$. In other words, the
CP derived under the single-antenna regime could serve as the lower
bound of that under the multi-antenna regime. According to Theorem
\ref{theorem: CP and ST scaling law SISO} in Section \ref{sec:Preliminary Analysis},
we can that $\mathsf{CP}_{N-\mathrm{L}}^{\mathrm{M}}\left(\lambda\right)\sim e^{-\bar{\kappa}\lambda}$.

Next, we analyze the CP upper bound as follows. Similarly as (\ref{eq:scaling MISO proof 1}),
we have
\begin{align}
\mathsf{CP}_{N}^{\mathrm{M}}\left(\lambda\right)= & \mathbb{P}\left\{ \left\Vert \mathbf{h}_{\mathrm{U}_{0},\mathrm{BS}_{0}}\mathbf{v}_{\mathrm{U}_{0},\mathrm{BS}_{0}}^{\mathrm{T}}\right\Vert ^{2}>s_{N}^{\mathrm{S}}I_{\mathrm{IC}}\right\} =\mathbb{P}\left\{ \stackrel[i=1]{N_{\mathrm{a}}}{\sum}g_{i}>s_{N}^{\mathrm{S}}I_{\mathrm{IC}}\right\} \nonumber \\
\stackrel[\left(\mathrm{a}\right)]{\lambda\rightarrow\infty}{<} & \mathbb{P}\left\{ g_{i}>\frac{s_{N}^{\mathrm{S}}I_{\mathrm{IC}}}{N_{\mathrm{a}}}\right\} =\mathsf{CP}_{N-\mathrm{U}}^{\mathrm{M}}\left(\lambda\right),\label{eq:scaling MISO proof 2}
\end{align}
where $g_{i}$ $\left(i=1,2,\ldots N_{\mathrm{a}}\right)$ follows
independently exponential distribution with mean $\frac{1}{2}$, i.e.,
$g_{i}\sim\mathrm{Exp}\left(\frac{1}{2}\right)$. Then, we explain
the reason why the inequality (a) holds true. Given $s_{N}^{\mathrm{S}}I_{\mathrm{IC}}$,
it is straightforward to obtain $\mathbb{P}\left\{ \left\Vert \mathbf{h}_{\mathrm{U}_{0},\mathrm{BS}_{0}}\mathbf{v}_{\mathrm{U}_{0},\mathrm{BS}_{0}}^{\mathrm{T}}\right\Vert ^{2}>s_{N}^{\mathrm{S}}I_{\mathrm{IC}}\right\} =\frac{\Gamma\left(N_{\mathrm{a}},\frac{s_{N}^{\mathrm{S}}I_{\mathrm{IC}}}{2}\right)}{\Gamma\left(N_{\mathrm{a}}\right)}$
and $\mathbb{P}\left\{ g_{i}>\frac{s_{N}^{\mathrm{S}}I_{\mathrm{IC}}}{N_{\mathrm{a}}}\right\} =\exp\left(-\frac{s_{N}^{\mathrm{S}}I_{\mathrm{IC}}}{2N_{\mathrm{a}}}\right)$.
Accordingly, it is easy to show $\mathbb{P}\left\{ \left\Vert \mathbf{h}_{\mathrm{U}_{0},\mathrm{BS}_{0}}\mathbf{v}_{\mathrm{U}_{0},\mathrm{BS}_{0}}^{\mathrm{T}}\right\Vert ^{2}>s_{N}^{\mathrm{S}}I_{\mathrm{IC}}\right\} <\mathbb{P}\left\{ g_{i}>\frac{s_{N}^{\mathrm{S}}I_{\mathrm{IC}}}{N_{\mathrm{a}}}\right\} $
when $s_{N}^{\mathrm{S}}I_{\mathrm{IC}}$ is sufficiently large or
equivalently $\lambda$ is sufficiently large. Hence, $\exists m>0$,
$\lambda_{0}$, $\forall\lambda>\lambda_{0}$, $\left|\mathsf{CP}_{N}^{\mathrm{M}}\left(\lambda\right)\right|\leq m\left|\mathsf{CP}_{N-\mathrm{U}}^{\mathrm{M}}\left(\lambda\right)\right|$.
According to Definition \ref{definition: scaling behavior}, we have
$\mathsf{CP}_{N}^{\mathrm{M}}\left(\lambda\right)=\mathcal{O}\left(\mathsf{CP}_{N-\mathrm{U}}^{\mathrm{M}}\left(\lambda\right)\right)$.

Following Theorem \ref{theorem: CP and ST scaling law SISO} in Section
\ref{sec:Preliminary Analysis}, it can be easily shown that $\mathsf{CP}_{N-\mathrm{U}}^{\mathrm{M}}\left(\lambda\right)\sim e^{-\bar{\kappa}\lambda}$.

Based on (\ref{eq:scaling MISO proof 1}) and (\ref{eq:scaling MISO proof 2}),
$\mathsf{CP}_{N}^{\mathrm{M}}\left(\lambda\right)\sim\lambda e^{-\bar{\kappa}\lambda}$
and $\mathsf{ST}_{N}^{\mathrm{M}}\left(\lambda\right)\sim\lambda e^{-\bar{\kappa}\lambda}$
hold true.

\subsection{Proof for Proposition \ref{proposition: CP and ST MISO approximation}\label{subsec:Proof for MISO APP}}

As indicated by Appendix \ref{subsec:Proof for Corollary CP and ST MISO},
the complicated form of CP is mainly due to $\left\Vert \mathbf{h}_{\mathrm{U}_{0},\mathrm{BS}_{0}}\mathbf{v}_{\mathrm{U}_{0},\mathrm{BS}_{0}}^{\mathrm{T}}\right\Vert ^{2}\sim\chi_{2N_{\mathrm{a}}}^{2}$.
To derive an approximate expression of CP instead, we propose to use
an exponentially distributed random variable $\tilde{g}_{0}$ with
mean $2N_{\mathrm{a}}$, i.e., $\tilde{g}_{0}\sim\mathrm{Exp}\left(\frac{1}{2N_{\mathrm{a}}}\right)$,
to approximate $\left\Vert \mathbf{h}_{\mathrm{U}_{0},\mathrm{BS}_{0}}\mathbf{v}_{\mathrm{U}_{0},\mathrm{BS}_{0}}^{\mathrm{T}}\right\Vert ^{2}$.
In consequence, $\mathsf{CP}_{N}^{\mathrm{M}}\left(\lambda\right)$
could be approximated by $\tilde{\mathsf{CP}}_{N}^{\mathrm{M}}\left(\lambda\right)$
given by
\begin{align}
\tilde{\mathsf{CP}}_{N}^{\mathrm{M}}\left(\lambda\right)= & \mathbb{P}\left\{ \tilde{g}_{0}>s_{N}^{\mathrm{S}}I_{\mathrm{IC}}\right\} =\mathbb{E}_{d_{0},\tilde{\Pi}_{\mathrm{BS}}}\left[\underset{\tiny{\mathrm{BS}_{i}\in\tilde{\Pi}_{\mathrm{BS}}}}{\prod}\frac{1}{1+\frac{s_{N}^{\mathrm{S}}}{N_{\mathrm{a}}}Pl_{N}\left(d_{i}\right)}\right]\nonumber \\
= & \mathbb{E}_{d_{0}}\left[\exp\left(-2\pi\lambda\int_{d_{0}}^{\infty}x\left(1-\frac{1}{1+\frac{s_{N}^{\mathrm{S}}}{N_{\mathrm{a}}}Pl_{N}\left(x\right)}\right)\mathrm{d}x\right)\right].\label{eq:CP and ST MISO approximation 1}
\end{align}
where $s_{N}^{\mathrm{S}}=\frac{\tau}{Pl_{N}\left(d_{0}\right)}$.
The remaining of the proof can be completed by following Appendix
\ref{subsec:Proof for SP and ST SISO} and thus is omitted.

\subsection{Proof for Theorem \ref{theorem: feasible region MSPM}\label{subsec:Proof for Theorem feasible region MSPM}}

It is shown from Fig. \ref{fig:CP scaling law MISO - approximation}
that the approximate CP in Proposition \ref{proposition: CP and ST MISO approximation}
could serve as a lower bound of the exact CP given in Corollary \ref{corollary: CP and ST MISO}.
Therefore, it is valid to use the approximate CP, which is in simple
form, to derive the necessary condition.

Following Theorem \ref{theorem: CP and ST scaling law MISO}, CP is
a decreasing function of $\lambda$. In other words, the maximal CP
is obtained when $\lambda\rightarrow0$ in the interference-limited
regime. According to Appendices \ref{subsec:Proof for scaling law}
and \ref{subsec:Proof for scaling law MISO}, the approximate CP under
SU-BF is given by $\mathsf{CP}_{N}^{\mathrm{M}}\left(\lambda\right)$
\begin{equation}
=\mathbb{E}_{r_{0}\in\left[R_{0},R_{N-1}\right)}\left[e^{-2\pi\lambda\int_{d_{0}}^{\infty}x\left(1-\frac{1}{1+s_{N}^{\mathrm{M}}Pl_{N}\left(x\right)}\right)\mathrm{d}x}\right]+\mathbb{E}_{r_{0}\in\left[R_{N-1},R_{N}\right)}\left[e^{-2\pi\lambda\int_{d_{0}}^{\infty}x\left(1-\frac{1}{1+s_{N}^{\mathrm{M}}Pl_{N}\left(x\right)}\right)\mathrm{d}x}\right],\label{eq:proof for theorem condition 0}
\end{equation}
where $s_{N}^{\mathrm{M}}=\frac{\tau^{\dagger}}{Pl_{N}\left(d_{0}\right)}$
and $\tau^{\dagger}=\frac{\tau}{N_{\mathrm{a}}}$. When $\lambda\rightarrow0$,
the CP of the typical downlink user $\mathrm{U}_{0}$ is dominated
by the interfering BSs, which are located within $\left(R_{N-1},R_{N}\right)$.
Note that $R_{N}=\infty$. Therefore, the maximal CP is given by
\begin{align}
\mathsf{CP}_{N-\mathrm{max}}^{\mathrm{M}}\left(\lambda\right)\overset{\left(\mathrm{a}\right)}{=} & \mathbb{E}_{r_{0}\in\left[R_{N-1},R_{N}\right)}\left[e^{-2\pi\lambda\int_{d_{0}}^{\infty}x\left(1-\frac{1}{1+s_{N}^{\mathrm{M}}Pl_{N}\left(x\right)}\right)\mathrm{d}x}\right]=\frac{e^{-\pi\lambda\left[R_{N-1}^{2}+\delta\left(\tau^{\dagger},\alpha_{N-1}\right)\left(R_{N-1}^{2}+\Delta h^{2}\right)\right]}}{1+\delta\left(\tau^{\dagger},\alpha_{N-1}\right)}\nonumber \\
\overset{\lambda\rightarrow0}{=} & \frac{1}{1+\delta\left(\tau^{\dagger},\alpha_{N-1}\right)},\label{eq:proof for theorem condition 1}
\end{align}
where $\delta\left(\tau^{\dagger},\alpha_{N-1}\right)=\frac{2\tau^{\dagger}\omega_{1}\left(\tau^{\dagger},\alpha_{N-1}\right)}{\alpha_{N-1}-2}$
and (a) follows because
\begin{align*}
\mathbb{E}_{r_{0}\in\left[R_{0},R_{N-1}\right)}\left[e^{-2\pi\lambda\int_{d_{0}}^{\infty}x\left(1-\frac{1}{1+s_{N}^{\mathrm{M}}Pl_{N}\left(x\right)}\right)\mathrm{d}x}\right]\overset{\lambda\rightarrow0}{=} & 0.
\end{align*}
Therefore, the necessary condition to meet the user CP requirement
can be obtained by solving
\begin{align}
\frac{1}{1+\delta\left(\tau^{\dagger},\alpha_{N-1}\right)}> & \varepsilon.\label{eq:proof for theorem condition 2}
\end{align}

\subsection{Proof for Corollary \ref{corollary: critical density SSPM}\label{subsec:Proof for critical density SSPM}}

It is shown from Theorem \ref{theorem: CP and ST scaling law SISO}
that $\mathsf{ST}_{N}^{\mathrm{M}}\left(\lambda\right)$ is a concave
function of $\lambda$. Hence, without the CP requirement, it is straightforward
to obtain $\lambda_{1}^{\dagger}$ by solving $\frac{\partial\mathsf{ST}_{1}^{\mathrm{M}}\left(\lambda\right)}{\partial\lambda}=0$,
where $\mathsf{ST}_{1}^{\mathrm{M}}\left(\lambda\right)$ is given
by Proposition \ref{proposition: CP and ST MISO approximation}. With
the CP requirement $\varepsilon$, the critical density $\lambda_{1}^{*}$
is given by $\lambda_{1}^{\dagger}$ (when $\varepsilon$ is small)
or by solving $\mathsf{CP}_{1}^{\mathrm{M}}\left(\lambda\right)=\varepsilon$
(when $\varepsilon$ is large), where $\mathsf{CP}_{1}^{\mathrm{M}}\left(\lambda\right)$
is given by Proposition \ref{proposition: CP and ST MISO approximation}.
Hence, the proof is complete.

\subsection{Proof for Corollary \ref{corollary: critical density DSPM}\label{subsec:Proof for critical density DSPM}}

When DSPM serves as the pathloss model, according to (\ref{eq:proof for theorem condition 0}),
we have
\begin{align}
\tilde{\mathsf{CP}}_{2}^{\mathrm{M}}\left(\lambda\right)= & \mathbb{E}_{r_{0}\in\left[0,R_{1}\right)}\left[e^{-2\pi\lambda\int_{d_{0}}^{\infty}x\left(1-\frac{1}{1+s_{2}^{\mathrm{M}}Pl_{2}\left(x\right)}\right)\mathrm{d}x}\right]+\mathbb{E}_{r_{0}\in\left[R_{1},\infty\right)}\left[e^{-2\pi\lambda\int_{d_{0}}^{\infty}x\left(1-\frac{1}{1+s_{2}^{\mathrm{M}}Pl_{2}\left(x\right)}\right)\mathrm{d}x}\right],\label{eq:critical density DSPM 1}
\end{align}
where $s_{2}^{\mathrm{M}}=\frac{\tau^{\dagger}}{Pl_{2}\left(d_{0}\right)}.$
Given $r_{0}\in\left[0,R_{1}\right)$, $s_{2}^{\mathrm{M}}=\frac{\tau^{\dagger}}{Pd_{0}^{\alpha_{0}}}$
and we have $\int_{d_{0}}^{\infty}x\left(1-\frac{1}{1+\frac{\tau^{\dagger}}{d_{0}^{\alpha_{0}}}l_{2}\left(x\right)}\right)\mathrm{d}x$
\begin{align}
= & \int_{d_{0}}^{R_{1}}x\left(1-\frac{1}{1+\frac{\tau^{\dagger}}{d_{0}^{\alpha_{0}}}x^{-\alpha_{0}}}\right)\mathrm{d}x+\int_{R_{1}}^{\infty}x\left(1-\frac{1}{1+\frac{\tau^{\dagger}}{d_{0}^{\alpha_{0}}}K_{1}x^{-\alpha_{1}}}\right)\mathrm{d}x\label{eq:critical density DSPM 2}\\
\overset{\left(\mathrm{a}\right)}{>} & \int_{d_{0}}^{\infty}x\left(1-\frac{1}{1+\frac{\tau^{\dagger}}{d_{0}^{\alpha_{0}}}x^{-\alpha_{0}}}\right)\mathrm{d}x\label{eq:critical density DSPM 3}\\
= & d_{0}^{2}\delta\left(\tau^{\dagger},\alpha_{0}\right),\label{eq:critical density DSPM 4}
\end{align}
where (a) follows because we use $x^{-\alpha_{0}}$ to replace $K_{1}x^{-\alpha_{1}}$
in the second term of (\ref{eq:critical density DSPM 2}). Equivalently,
the interference power is strengthened and the inequality in (\ref{eq:critical density DSPM 3})
holds. Given $r_{0}\in\left[R_{1},\infty\right)$, $s_{2}^{\mathrm{M}}=\frac{\tau^{\dagger}}{PK_{1}d_{0}^{-\alpha_{1}}}$
and we have
\begin{equation}
\int_{d_{0}}^{\infty}x\left(1-\frac{1}{1+s_{2}^{\mathrm{M}}Pl_{2}\left(x\right)}\right)\mathrm{d}x=\int_{d_{0}}^{\infty}x\left(1-\frac{1}{1+\frac{\tau^{\dagger}}{d_{0}^{-\alpha_{1}}}x^{-\alpha_{1}}}\right)\mathrm{d}x=d_{0}^{2}\delta\left(\tau^{\dagger},\alpha_{1}\right).\label{eq:critical density DSPM 5}
\end{equation}
Substituting (\ref{eq:critical density DSPM 4}) and (\ref{eq:critical density DSPM 5})
into (\ref{eq:critical density DSPM 1}),
\begin{align}
\tilde{\mathsf{CP}}_{2}^{\mathrm{M}}\left(\lambda\right)= & \frac{1-\exp\left[-\pi\lambda\left(R_{1}^{2}\left(1+\delta\left(\tau^{\dagger},\alpha_{0}\right)\right)+\bigtriangleup h^{2}\delta\left(\tau^{\dagger},\alpha_{0}\right)\right)\right]}{1+\delta\left(\tau^{\dagger},\alpha_{0}\right)}\nonumber \\
+ & \frac{\exp\left[-\pi\lambda\left(R_{1}^{2}\left(1+\delta\left(\tau^{\dagger},\alpha_{1}\right)\right)+\bigtriangleup h^{2}\delta\left(\tau^{\dagger},\alpha_{1}\right)\right)\right]}{1+\delta\left(\tau^{\dagger},\alpha_{1}\right)}.\label{eq:critical density DSPM 6}
\end{align}
When $R_{1}$ is large, it is easy to show that the first term in
(\ref{eq:critical density DSPM 6}) is much smaller than the second
term. Therefore, we directly use the second term as a substitution
of $\tilde{\mathsf{CP}}_{2}^{\mathrm{M}}\left(\lambda\right)$. The
remaining of the proof can be completed according to the proof for
Corollary \ref{corollary: critical density SSPM} in Appendix \ref{subsec:Proof for critical density SSPM}
and thus omitted.

\bibliographystyle{IEEEtran}
\bibliography{ref_BPM}

\end{document}